\documentclass[showpacs,preprintnumbers,nofootinbib,amsmath,amssymb]{revtex4}
\usepackage{hyperref,slashed,color}
\usepackage{graphicx,subfig}
\allowdisplaybreaks

   \def\sk{\Sigma_k}

  \def\fm{f_-}
   \def\fp{f_+}

   \def\chip{\chi_+}
    \def\chim{\chi_-}
     \def\la{\langle}
      \def\ra{\rangle}
 \def\skp{\Sigma_k'}
 \def\skpp{\Sigma_k''}
 \def\skppp{\Sigma_k'''}

  \def\otw{\widetilde{O}^W}
  
\begin{document}
\preprint{TUHEP-TH-09170}
\title{Computation of the  coefficients for  the order $p^6$ anomalous chiral Lagrangian}

\bigskip
\author{Shao-Zhou
Jiang$^{1,2}$\footnote{Email:\href{mailto:jsz@mails.tsinghua.edu.cn}{jsz@mails.tsinghua.edu.cn}.},
 and Qing Wang$^{1,2}$\footnote{Email:
\href{mailto:wangq@mail.tsinghua.edu.cn}{wangq@mail.tsinghua.edu.cn}.}\footnote{corresponding
author}\\~}

\bigskip
\affiliation{$^1$Center for High Energy Physics, Tsinghua University, Beijing 100084, P.R. China\\
$^2$Department of Physics, Tsinghua University, Beijing 100084, P.R.
China\footnote{mailing address}}

\begin{abstract}
We present the results of computing the order $p^6$ low energy
constants in the anomalous part of the chiral Lagrangian for both
two and three flavor pseudoscalar mesons. This is a generalization
of our previous work on calculating the order $p^6$
 coefficients for the normal part of the chiral Lagrangian in
 terms of the quark self energy
$\Sigma(p^2)$. We show that most of our results are consistent with
those we have found in the literature.
\end{abstract}
\pacs{12.39.Fe, 11.30.Rd, 12.38.Aw, 12.38.Lg} \maketitle
\section{Introduction and Background}

It is well known that the chiral symmetry in quantum chromodynamics
(QCD) suffers anomalies due to the non-invariance of the path
integral measure of the quark fields under the chiral symmetry
transformation. The anomaly reflects the fact that the classical
chiral symmetry may be violated by quantum corrections. At the level
of the effective chiral Lagrangian for the pseudoscalar meson field
$U$, anomaly no longer comes from the path integral measure. Instead
it is due to the non-invariance of the effective chiral Lagrangian.
If we denote by $\Gamma_\mathrm{eff}[U,J]$ the effective action for
the pseudoscalar meson field $U$ and the external source $J$, then
this non-invariance can be expressed as
\begin{eqnarray}
\Gamma_\mathrm{eff}[U,J]-\Gamma_\mathrm{eff}[U_\Omega,J_\Omega]=\Gamma[\Omega,J]\;,\label{EffAction1}
\end{eqnarray}
where $U_\Omega\equiv\Omega^{\dag}U\Omega^{\dag}$ and
$J_\Omega\equiv[\Omega
P_R+\Omega^{\dag}P_L][J+\slashed{\partial}][\Omega P_R+\Omega^\dag
P_L]$.  $\Gamma[\Omega,J]$ is the anomaly from the light quark path
integral measure
$\mathcal{D}\bar{\psi}_\Omega\mathcal{D}\psi_\Omega=
\mathcal{D}\bar{\psi}\mathcal{D}\psi~e^{\Gamma}$ or the well known
Wess-Zumino-Witten term. We can formally express it as
\begin{eqnarray}
\Gamma[\Omega,J]&=&-\ln\mathrm{Det}[(\Omega
P_R+\Omega^{\dag}P_L)(\Omega
P_R+\Omega^{\dag}P_L)]=-\mathrm{Tr}\ln[\slashed{\partial}+J_{\Omega}]+\mathrm{Tr}\ln[\slashed{\partial}+J]\;,~~~~\label{QCDlevel}
\end{eqnarray}
 Because
for $N_f$ light quarks, each generator of the chiral symmetry
$SU(N_f)_L\otimes SU(N_f)_R/SU(N_f)_V$ corresponds to a Goldstone
boson, which is treated phenomenologically as the physical
pseudoscalar meson field,  the phase angle of the chiral rotation
group element $\Omega$ can be treated as the pseudoscalar meson
field, i.e. $U=\Omega^2$. Then comparing (\ref{EffAction1}) and
(\ref{QCDlevel}), we can rewrite the effective action
$\Gamma_\mathrm{eff}$ as
\begin{eqnarray}
\Gamma_\mathrm{eff}[U,J]=-\mathrm{Tr}\ln[\slashed{\partial}+J_{\Omega}]+\mathrm{Tr}\ln[\slashed{\partial}+J]+F[U,J]
\hspace*{2cm}F[U,J]=F[U_\Omega,J_\Omega] \label{AnomalyApp}
\end{eqnarray}
The $U$ and $J$ dependence for $F[U,J]$ is not fixed by
(\ref{EffAction1}), but $F[U,J]$ is invariant on $U\rightarrow
U_\Omega$ and $J\rightarrow J_\Omega$. Hence $F[U,J]$ represents
those chiral invariant terms. In fact
 $U_\Omega=\Omega^\dag\Omega^2\Omega^\dag=1$, and
$\Gamma_\mathrm{eff}[U_\Omega,J_\Omega]
=\Gamma_\mathrm{eff}[1,J_\Omega]
=-\mathrm{Tr}\ln[\slashed{\partial}+J_\Omega]+\mathrm{Tr}\ln[\slashed{\partial}+J_\Omega]+F[U_\Omega,J_\Omega]=
F[U,J]$. Note that the effective action is the path integration
result for $S_\mathrm{eff}[U,J]$, the action of the effective chiral
Lagrangian for the pseudoscalar meson field $U$ and the external
source $J$,
\begin{eqnarray}
e^{-\Gamma_\mathrm{eff}[U_{cl},J]}=\int\mathcal{D}U~e^{-S_\mathrm{eff}[U,J]}
\hspace{2cm}U_{cl}(x)\equiv\int\mathcal{D}U~U(x)~e^{-S_\mathrm{eff}[U,J]}\;,
\label{EffAction}
\end{eqnarray}
where the second equation gives the definition of $U_{cl}$ which
fixes
 $U_{cl}$ as the functional of the external source $J$. With
(\ref{AnomalyApp}), (\ref{EffAction}) becomes
\begin{eqnarray}
e^{\mathrm{Tr}\ln[\slashed{\partial}+J_{\Omega}]-\mathrm{Tr}\ln[\slashed{\partial}+J]-F(U_{cl},J)}=\int\mathcal{D}U~e^{-S_\mathrm{eff}[U,J]}\;.
\label{EffAction2}
\end{eqnarray}
Ref.\cite{AnomApproach},\cite{AnomApproach1},\cite{AnomApproach2}
choose as an approximation
\begin{eqnarray}
S_\mathrm{eff,0}[U,J]=-\mathrm{Tr}\ln[\slashed{\partial}+J_{\Omega}]+\mathrm{Tr}\ln[\slashed{\partial}+J]\;,\label{AnomalyApp1}
\end{eqnarray}
where subscript $_0$ is used to denote the approximation. From
(\ref{EffAction1}), (\ref{AnomalyApp}) and (\ref{EffAction2}), we
find that under the chiral symmetry transformation,
$S_\mathrm{eff,0}[U,J]$, defined in (\ref{AnomalyApp1}), is not
invariant. Substituting (\ref{AnomalyApp1}) back into
(\ref{EffAction2}) and using standard loop expansion as developed in
Ref.\cite{LoopExp}, we find $F[U_{cl},J]$ is the pure loop
correction from the action $S_\mathrm{eff,0}[U,J]$. From the action
(\ref{AnomalyApp1}), one can calculate various low energy constants
(LECs) of the effective chiral Lagrangian for pseudoscalar mesons.
In Ref.\cite{WQ3}, we call (\ref{AnomalyApp1}) the anomaly approach.
 In our previous paper \cite{WQ4}, we have shown that the
finite order $p^4$ LECs of the normal part of
$S_\mathrm{eff,0}[U,J]$ are exactly canceled by the summation of all
the $p^6$ and higher order terms. Eq.(\ref{QCDlevel}) further shows
that even for the anomalous part, $S_\mathrm{eff,0}[U,J]$ only
contributes the Wess-Zumino-Witten term; it cannot produce the $p^6$
and higher order anomaly terms. This absence of the normal part and
the $p^6$ and more higher order anomalous part reflects the fact
that the choice of (\ref{AnomalyApp1}) is not correct, although it
offers the correct Wess-Zumino-Witten term. Further,
(\ref{AnomalyApp1}) is independent of the strong interaction
dynamics, i.e., even we switch off the quark-gluon interaction by
deleting the strong interaction coupling constant,
(\ref{AnomalyApp1}) is not changed. These facts imply that we need
to add  some strong dynamics dependent correction term $\Delta
S_\mathrm{eff}[U,J]$ to $S_\mathrm{eff,0}[U,J]$ as given in
(\ref{AnomalyApp1}),
\begin{eqnarray}
S_\mathrm{eff}[U,J]=S_\mathrm{eff,0}[U,J]+\Delta
S_\mathrm{eff}[U,J]\;.\label{Seff}
\end{eqnarray}
From (\ref{EffAction2}) and (\ref{AnomalyApp1}), we find that
$\Delta S_\mathrm{eff}[U,J]$, introduced in (\ref{Seff}), must be
invariant under chiral symmetry transformations. In Refs.\cite{WQ1}
and \cite{WQ2}, $\Delta S_\mathrm{eff}[U,J]$ is taken to be
\begin{eqnarray}
\Delta
S_\mathrm{eff}[U,J]=\mathrm{Tr}\ln[\slashed{\partial}+J_{\Omega}+\Sigma(-\bar{\nabla}^2)]
\label{DeltaSeff}
\end{eqnarray}
with $\Sigma$ being the quark self energy satisfying the
Schwinger-Dyson equation (SDE) and $\bar{\nabla}^\mu$ is defined as
$\bar{\nabla}^\mu\equiv\partial^\mu-iv^\mu_\Omega$.
 This expression for $\Delta S_\mathrm{eff}[U,J]$ encodes the dynamics of
the underlying QCD through quark self energy $\Sigma$ and in
Ref.\cite{WQma}, we have shown that (\ref{DeltaSeff}) does not
produce the Wess-Zumino-Witten term ensuring the correctness of
(\ref{EffAction1}).

 In Ref.\cite{WQ1}, we have calculated the orders $p^2$ and $p^4$ normal part LECs in terms of
the action (\ref{Seff}) and (\ref{DeltaSeff}). The importance of
knowledge of LECs of the chiral Lagrangian,
 especially for order $p^6$ LECs  was emphasized in
Ref.\cite{Review}. Recently, in Ref.\cite{WQ4}, we improved the
computation procedure and generalized the calculations up to
 the order $p^6$ normal part LECs. In Ref.\cite{WQma}, we have
  calculated the $p^4$ order anomalous part and shown that the $\Sigma$
 dependent coefficient generates the correct coefficient $N_c$ for the Wess-Zumino-Witten
 term. It is the purpose of this paper to calculate all order $p^6$
 LECs for the anomalous part of the chiral Lagrangian (\ref{Seff}). In
 fact the general structure of the $p^6$ order anomalous part chiral
 Lagrangian was first given by Refs.\cite{anom-1} and \cite{anom0}
 and later clarified by Refs.\cite{anom1} and \cite{anom2}.
  Ref.\cite{p6anomLEC} estimates the values of several of the order $p^6$ LECs for the anomalous part of the chiral Lagrangian. Although
  order $p^6$ LECs for the normal part of the chiral
 Lagrangian seem attract more attentions in the literature (see references given in
 \cite{WQ4}), they are the next next to leading order terms. The order $p^6$ LECs for the anomalous part of the
 chiral Lagrangian are belong to the next leading order terms.

 This paper is organized as follows: in Sec.II, we review the calculation of the order $p^4$
 anomalous part of the chiral Lagrangian in terms of the
action
 (\ref{Seff}).  With the method used in section II, in Sec.III, we
 compute the order $p^6$ LECs  for the anomalous part of the chiral Lagrangian,
 and obtain the analytical expression for the LECs
 in terms of quark self energy $\Sigma$. We further compute the numerical values for these
 LECs.
We compare our results with those obtained in
 literature. Sec.IV is the summary and future directions of our work. We list some necessary tables and formulae in
 appendices.
\section{Review the
 order $p^4$ anomalous part of the chiral Lagrangian}

For the anomalous part of the chiral Lagrangian, the leading
nontrivial order is $p^4$ and it is the well known
Wess-Zumino-Witten term. In Ref.\cite{WQma}, we have calculated the
action (\ref{Seff}) by several different methods and all obtain the
same Wess-Zumino-Witten term. If we naively apply these methods to
the next to leading order $p^6$ computations, we will find that they
are too complex to be achieved even with the help of the computer.
In this section, we build a method which is suitable to be
generalized to the order $p^6$ calculations. The order $p^4$ of the
anomalous chiral Lagrangian is here only to be used to explain our
method. Ref.\cite{WQma} only expresses the Wess-Zumino-Witten term
in terms of a parameter integration. In this section, we will
explicitly finish this parameter integration and show that it does
recover the Wess-Zumino-Witten term.

Since we are only interested in the $U$ field dependent part of the
anomalous part of the chiral Lagrangian, we can drop out the pure
source terms. Then our choice of $\Delta S_\mathrm{eff}[U,J]$ in
(\ref{DeltaSeff}) gives the result that only $\Sigma$ dependent
terms in $\Delta S_\mathrm{eff}[U,J]$ contribute to the chiral
Lagrangian, while the $\Sigma$ independent terms in $\Delta
S_\mathrm{eff}[U,J]$ are completely canceled by the term
$-\mathrm{Tr}\ln[\slashed{\partial}+J_\Omega]$ in
$S_\mathrm{eff,0}[U,J]$, leaving a pure $U$ field independent term
$\mathrm{Tr}\ln[\slashed{\partial}+J]$. So what we need to compute
is
\begin{eqnarray}
S_\mathrm{eff}[U,J]=\bigg[\mathrm{Tr}\ln[\slashed{\partial}+J_{\Omega}+\Sigma(-\bar{\nabla}^2)]
-\mathrm{Tr}\ln[\slashed{\partial}+J+\Sigma(-\nabla^2)]\bigg]_{\Sigma~\mbox{\tiny
dependent}}\;,\label{Seff1}
\end{eqnarray}
in which we have added in $S_\mathrm{eff}[U,J]$ an extra pure source
term
$-\mathrm{Tr}\ln[\slashed{\partial}+J+\Sigma(-\nabla^2)]\bigg|_{\Sigma~\mbox{\tiny
dependent}}$ for later use, and we define
$\nabla^\mu\equiv\partial^\mu-iv^\mu$. Now we write $\Omega$ as
$\Omega=e^{-i\beta}$ and further introduce a parameter $t$ dependent
rotation element  $\Omega(t)=e^{-it\beta}$. With the help of the
relation $\Omega(1)=\Omega$ and $\Omega(0)=1$, (\ref{Seff1}) becomes
\begin{eqnarray}
S_\mathrm{eff}[U,J]=\mathrm{Tr}\ln[\slashed{\partial}+J_{\Omega(t)}+\Sigma(-\nabla_t^2)]
\bigg|^{t=1}_{t=0,~\Sigma~\mbox{\tiny
dependent}}\hspace{2cm}\nabla_t^\mu=\overline{\nabla}^\mu\bigg|_{\Omega\rightarrow\Omega(t)}\label{Sdifference}
\end{eqnarray}
with $\nabla^\mu_t=\partial^\mu-iv^\mu_{\Omega(t)}$. $J_{\Omega(t)}$
is $J_\Omega$ with $\Omega$ replaced by $\Omega(t)$. We decompose
$J_\Omega$ as
$J_\Omega=-i\slashed{v}_\Omega-i\slashed{a}_\Omega\gamma_5-s_\Omega+ip_\Omega\gamma_5$,
so we can also decompose $J_{\Omega(t)}$ as
$J_{\Omega(t)}=-i\slashed{v}_t-i\slashed{a}_t\gamma_5-s_t+ip_t\gamma_5$.
Result (\ref{Sdifference}) implies that our chiral Lagrangian can be
expressed as the difference of Trln($\cdots$) at $t$ dependent
chiral rotation between $t=1$ and $t=0$. Since the $t$ dependent
rotated source $J_{\Omega(t)}$ satisfies
\begin{eqnarray}
\frac{\partial J_{\Omega(t)}}{\partial t}=\frac{1}{2}[\frac{\partial
U_t}{\partial
t}U^\dag_t\gamma_5,\slashed{\partial}+J_{\Omega(t)}]_+\hspace{2cm}U_t=\Omega^2(t)\;,
\end{eqnarray}
we can further proceed to express the chiral Lagrangian in terms of
integration over the parameter $t$:
\begin{eqnarray}
S_\mathrm{eff}[U,J]&=&\int_0^1dt~\frac{d}{dt}\mathrm{Tr}\ln[i\slashed{\partial}+J_{\Omega(t)}+\Sigma(-\nabla_t^2)]
\bigg|_{\Sigma~\mbox{\tiny dependent}}\nonumber\\
&=& \int_0^1dt~\mathrm{Tr}\bigg[[\frac{\partial
J_{\Omega(t)}}{\partial
t}+\frac{\partial\Sigma(-\nabla_t^2)}{\partial
t}][i\slashed{\partial}+J_{\Omega(t)}+\Sigma(-\nabla_t^2)]^{-1}\bigg]_{\Sigma~\mbox{\tiny
dependent}}\nonumber\\
&=& \int_0^1dt~\mathrm{Tr}\bigg[\bigg(\frac{1}{2}[\frac{\partial
U_t}{\partial
t}U^\dag_t\gamma_5,\slashed{\partial}+J_{\Omega(t)}]_++\frac{\partial\Sigma(-\nabla_t^2)}{\partial
t}\bigg)[i\slashed{\partial}+J_{\Omega(t)}+\Sigma(-\nabla_t^2)]^{-1}\bigg]_{\Sigma~\mbox{\tiny
dependent}}\;.\label{Seff2}
\end{eqnarray}
(\ref{Seff2}) is the main formula we rely on to calculate LECs.
Ref.\cite{WQma} explicitly calculates the order $p^4$ anomalous part
of the r.h.s. of (\ref{Seff2}) and finds the result
\begin{eqnarray}
S_\mathrm{eff}[U,J]\bigg|_{\mbox{\tiny
anomalous~}p^4}\hspace*{-1.3cm}&=&-2N_c\epsilon_{\mu\nu\alpha\beta}\int
d^4x\int_0^1dt\int\frac{d^4k}{(2\pi)^4}~\mathrm{tr}_f\bigg[\frac{\partial
U_t}{\partial
t}U^\dag_t\bigg(\frac{\Sigma(k^2)[\Sigma^2(k^2)-k^2][\Sigma(k^2)-2k^2\Sigma'(k^2)]}{[\Sigma^2(k^2)+k^2]^4}
\nonumber\\
&&\times(2\nabla^\mu_t\nabla^\nu_t\nabla^\alpha_t\nabla^\beta_t+2a^\mu_ta^\nu_t\nabla^\alpha_t\nabla^\beta_t
-2\nabla^\mu_ta^\nu_t\nabla^\alpha_ta^\beta_t+2\nabla^\mu_ta^\nu_ta^\alpha_t\nabla^\beta_t+2a^\mu\nabla^\nu_t\nabla^\alpha_ta^\beta_t-2a^\mu_t\nabla^\nu_ta^\alpha_t\nabla^\beta_t
\nonumber\\
&&+2\nabla^\mu_t\nabla^\nu_ta^\alpha_ta^\beta_t
+2a^\mu_ta^\nu_ta^\alpha_ta^\beta_t)+\frac{k^2\Sigma(k^2)[\Sigma(k^2)-2k^2\Sigma'(k^2)]}{[\Sigma^2(k^2)+k^2]^4}(4\nabla^\mu_t\nabla^\nu_t\nabla^\alpha_t\nabla^\beta_t
+2a^\mu_ta^\nu_t\nabla^\alpha_t\nabla^\beta_t
\nonumber\\
&&-2\nabla^\mu_ta^\nu_t\nabla^\alpha_ta^\beta_t+4a^\mu_t\nabla^\nu_t\nabla^\alpha_ta^\beta_t
-2a^\mu_t\nabla^\nu_ta^\alpha_t\nabla^\beta_t+2\nabla^\mu_t\nabla^\nu_ta^\alpha_ta^\beta_t)\bigg)\bigg]\;.
\end{eqnarray}
 The momentum integration can  be calculated analytically, because the
integrand is a total derivative. The result is
\begin{eqnarray}
S_\mathrm{eff}[U,J]\bigg|_{\mbox{\tiny
anomalous~}p^4}&=&\frac{1}{32\pi^2}\epsilon_{\mu\nu\alpha\beta}\int
d^4x\int_0^1dt~\mathrm{tr}_f\bigg[\frac{\partial U_t}{\partial
t}U^\dag_t~\bigg(V_t^{\mu\nu}V_t^{\alpha\beta}+\frac{2i}{3}[a^\mu_ta_t^\nu,V^{\alpha\beta}_t]_++
\frac{4}{3}d_t^{\mu}a^\nu_td_t^{\alpha}a^\beta_t\nonumber\\
&&+\frac{8i}{3}a^\mu_tV_t^{\nu\alpha}a_t^\beta+
\frac{4}{3}a^\mu_ta^\nu_ta^\alpha_ta^\beta_t\bigg)\bigg]\;,\label{Seff3}
\end{eqnarray}
where
$V^{\mu\nu}_t=\partial^{\mu}v^{\nu}_t-\partial^{\nu}v^{\mu}_t-i[v^{\mu}_t,v^{\nu}_t]$
and
$d^{\mu}_ta^{\nu}_t=\partial^{\mu}a^{\nu}_t-i[v^{\mu}_t,a^{\nu}_t]$.
 Ref.\cite{WQma} only gives the above result (\ref{Seff3}) without finishing the
integration over parameter $t$. Now we continue to achieve this
integration, with the help of following relations:
\begin{eqnarray}
&&\hspace{-0.3cm}\frac{\partial a_t^\mu}{\partial
t}=\frac{i}{2}[\nabla^\mu_t\frac{\partial U_t}{\partial
t}U^\dag_t-\frac{\partial U_t}{\partial
t}U^\dag_t\nabla^\mu_t]\hspace{4cm}\frac{\partial v_t^\mu}{\partial
t}=\frac{1}{2}[a^\mu_t\frac{\partial U_t}{\partial
t}U^\dag_t-\frac{\partial U_t}{\partial t}U^\dag_ta^\mu_t]\nonumber\\
&&\hspace{-0.3cm}\frac{\partial s_t}{\partial
t}=-\frac{i}{2}[p_t\frac{\partial U_t}{\partial
t}U^\dag_t+\frac{\partial U_t}{\partial
t}U^\dag_tp_t]\hspace{4.2cm}\frac{\partial p_t}{\partial
t}=\frac{i}{2}[s_t\frac{\partial U_t}{\partial
t}U^\dag_t+\frac{\partial U_t}{\partial t}U^\dag_ts_t]\label{difft}\\
&&\hspace{-0.3cm}\frac{\partial d^{\mu}a_t^\nu}{\partial
t}\!=\frac{i}{2}[(\nabla^\mu_t\nabla^\nu_t\!\!+\!a^\nu_ta^\mu_t)\frac{\partial
U_t}{\partial t}U^\dag_t\!-\!\nabla^\nu_t\frac{\partial
U_t}{\partial t}U^\dag_t\nabla^\mu_t\!-\!\nabla^\mu_t\frac{\partial
U_t}{\partial t}U^\dag_t\nabla^\nu_t\!-\!a^\nu_t\frac{\partial
U_t}{\partial t}U^\dag_ta^\mu_t\!-\!a^\mu_t\frac{\partial
U_t}{\partial t}U^\dag_ta^\nu_t\!+\!\frac{\partial U_t}{\partial
t}U^\dag_t(\nabla^\nu_t\nabla^\mu_t\!\!+\!a^\mu_ta^\nu_t)]~~~~\nonumber\\
&&\hspace{-0.3cm}\frac{\partial V^{\mu\nu}_t}{\partial
t}=\frac{1}{2}[-\nabla^\mu_t\frac{\partial U_t}{\partial
t}U^\dag_ta^\nu_t+\frac{\partial U_t}{\partial
t}U^\dag_t\nabla^\mu_ta^\nu_t-\frac{\partial U_t}{\partial
t}U^\dag_t d_t^{\mu}a^\nu_t+d_t^{\mu}a^\nu_t\frac{\partial
U_t}{\partial t}U^\dag_t +a^\nu_t\nabla^\mu_t\frac{\partial
U_t}{\partial t}U^\dag_t -a^\nu_t\frac{\partial U_t}{\partial
t}U^\dag_t\nabla^\mu_t\nonumber\\
&&\hspace{1cm}+\nabla^\nu_t\frac{\partial U_t}{\partial
t}U^\dag_ta^\mu_t -\frac{\partial U_t}{\partial
t}U^\dag_t\nabla^\nu_ta^\mu_t +\frac{\partial U_t}{\partial
t}U^\dag_t d_t^{\nu}a^\mu_t -d_t^{\nu}a^\mu_t\frac{\partial
U_t}{\partial t}U^\dag_t -a^\mu_t\nabla^\nu_t\frac{\partial
U_t}{\partial t}U^\dag_t +a^\mu_t\frac{\partial U_t}{\partial
t}U^\dag_t\nabla^\nu_t]\nonumber
\end{eqnarray}
and by lengthy calculations, we can rewrite (\ref{Seff3}) as
\begin{eqnarray}
S_\mathrm{eff}[U,J]\bigg|_{\mbox{\tiny anomalous~}p^4}&=&
-\frac{N_c}{48\pi^2}\int
d^4x\int_0^1dt~\epsilon_{\mu\nu\lambda\rho}\mathrm{tr}_f\bigg[\frac{\partial
U_t}{\partial t} U^\dag_t
R^\mu_tR^\nu_tR^\lambda_tR^\rho_t+\frac{d}{dt}W^{\mu\nu\lambda\rho}(U_t,l,r)\bigg]\label{WZW}
\end{eqnarray}
with $l^\mu=v^\mu-a^\mu,~r^\mu=v^\mu+a^\mu$,
$R^\mu_t=U^\dag_t\partial^{\mu}U_t$,
$L^\mu_t=(\partial^{\mu}U_t)U^\dag_t$ and
 \begin{eqnarray}
 W^{\mu\nu\lambda\rho}(U_t,l,r)&=&iR^{\mu}_tR^{\nu}_tR^{\lambda}_tl^{\rho}+l^{\mu}\partial^{\nu}l^{\lambda}R_t^{\rho}
 +\partial^{\mu}l^{\nu}l^{\lambda}R_t^{\rho}-\frac{1}{2}R_t^{\mu}l^{\nu}R_t^{\lambda}l^{\rho}
 +r^{\mu}U_tl^{\nu}R_t^{\lambda}R^{\rho}U_t^\dag+iR^\mu_tl^{\nu}l^{\lambda}l^{\rho}+iU^\dag_tr^{\mu}U_t\partial^{\nu}l^{\lambda}l^{\rho}
 \notag\\
 &&+iU^\dag_tr^{\mu}\partial^{\nu}r^{\lambda}U_tl^{\rho}-il^{\mu}U^\dag_tr^{\nu}U_tl^{\lambda}R^{\rho}_t
 -R^{\mu}_tU^{\dag}_t\partial^{\nu}r^{\lambda}U_tl^{\rho}
 +U^\dag_tr^{\mu}U_tl^{\nu}l^{\lambda}l^\rho
 +\frac{1}{4}U^\dag_tr^{\mu}U_tl^{\nu}U^\dag_tr^{\lambda}U_tl^{\rho}\notag\\
 &&-(U_t\leftrightarrow U^\dag_t,l^\mu\leftrightarrow
 r^\mu,L^\mu_t\leftrightarrow-R^{\mu}_t)\;.
 \end{eqnarray}
In Ref.\cite{WQma}, we already show that the first term of the
r.h.s. of Eq.(\ref{WZW}) is just the Wess-Zumino-Witten term of the
form defined on a four dimensional boundary disc $Q$ in five
dimensional space-time
\begin{eqnarray}
-\frac{N_c}{48\pi^2}\int
d^4x\int_0^1dt~\epsilon_{\mu\nu\lambda\rho}\mathrm{tr}_f\bigg[\frac{\partial
U_t}{\partial t} U^\dag_t R^\mu_tR^\nu_tR^\lambda_tR^\rho_t\bigg]
=-\frac{N_c}{240\pi^2}\int_Q
d\Sigma_{ijklm}\mathrm{tr}_f[R^iR^jR^kR^lR^m]
\end{eqnarray}
with $R^i\equiv U^\dag\partial^iU$. For the second term of the
r.h.s. of Eq.(\ref{WZW}), the integration over parameter $t$ can be
calculated explicitly,
\begin{eqnarray}
-\frac{N_c}{48\pi^2}\int
d^4x\int_0^1dt~\epsilon_{\mu\nu\lambda\rho}\mathrm{tr}_f\bigg[\frac{d}{dt}W^{\mu\nu\lambda\rho}(U_t,l,r)\bigg]
=-\frac{N_c}{48\pi^2}\int
d^4x~\epsilon_{\mu\nu\lambda\rho}\mathrm{tr}_f\bigg[W^{\mu\nu\lambda\rho}(U,l,r)-W^{\mu\nu\lambda\rho}(1,l,r)\bigg]\;,
\end{eqnarray}
which the just the gauge part of the Wess-Zumino-Witten term given
by Ref.\cite{anom1} and \cite{Zhou}. This finishes the explicit
calculation of the order $p^4$ anomalous part of the chiral
Lagrangian starting from formula (\ref{Seff2}). We leave  the order
$p^6$ part to the next section.
\section{Calculation of the order $p^6$ anomalous part of the chiral Lagrangian}

In this section, we start from Eq.(\ref{Seff2}) to calculate its
order $p^6$ anomalous part of the chiral Lagrangian. For
convenience, we change to the Minkowski space to perform our
calculations. Direct computation gives the result
\begin{eqnarray}
 S_\mathrm{eff}[U,J]\bigg|_{\mbox{\tiny
anomalous~}p^6}={\displaystyle\sum_{m=1}^{210}}\int
d^4x~\bar{K}_m^W\int_0^1dt~\mathrm{tr}_f
[\bar{O}_m^W(x,t)]\;,\label{p6-1}
\end{eqnarray}
where $\bar{K}_m^W$ is the coefficient in front of the operator
$\bar{O}_m^W(x,t)$, which depends on quark self energy
$\Sigma(k^2)$. The 210 parameter $t$ dependent operators
$\bar{O}_m^W(x,t)$ all have the structure of
$\bar{O}_m^W(x,t)=\epsilon_{\mu\nu\lambda\rho}\frac{\partial
U_t}{\partial t}U^\dag_t\bar{O}^{\mu\nu\lambda\rho}_m(x,t)$ and
$\bar{O}^{\mu\nu\lambda\rho}_m(x,t)$ are order $p^6$ operators
  consisting of multiplications of various compositions of $a^\mu_t$,
$\nabla^\nu_t$, $s_t$ and $p_t$. In Appendix A we list all these
operators in Table V. In obtaining (\ref{p6-1}), we have applied the
Schouten identity, which reduces the original total 294 operators to
the present 210 operators. In the literature,  the
general $p^6$ order anomalous part of the chiral Lagrangian given in
Ref.\cite{anom1} has only 24 independent
operators. For $N_f=3,2$ this number reduces to 23 and five respectively. Specially for the case of $N_f=2$,
 to incorporate the electro-magnetic field into the external
source $v^\mu$, the original traceless property of $v^\mu$ must be
dropped, this changes the original
five independent $p^6$ order anomalous operators into
 thirteen. If we denote the independent
operators
 by $O_n^W(x)$ ($o_n^W(x)$ for $N_f=2$) and corresponding coefficients in front of the
operators by  $C_n^W$ ($c_n^W(x)$ for $N_f=2$) respectively, then
 (\ref{p6-1}) becomes
 \begin{eqnarray}
 S_\mathrm{eff}[U,J]\bigg|_{\mbox{\tiny
anomalous~}p^6}={\displaystyle\sum_{n=1}^{24}}\int
d^4x~C_n^WO_n^W(x)\stackrel{N_f=2}{=====}{\displaystyle\sum_{n=1}^{13}}\int
d^4x~c_n^Wo_n^W(x)\;.\label{p6-2}
\end{eqnarray}
Note that our starting chiral Lagrangian (\ref{Seff}) only involves
one trace for flavor indices. If we further apply the equation of
motion
 to (\ref{p6-2}), there will appear some operators with two flavor
traces. Our result prohibits the appearance of three operators
$O_3^W, O_{18}^W, O_{24}^W$, leaving 21 independent operators. This
implies that our formulation gives $C_3^W=C_{18}^W=C_{24}^W=0$. If
we do not apply the equation of motion, there will be more
independent operators and now this number is 23.
 To make our calculation more convenient, we denote these operators before applying
the equation of motion
 by $\tilde{O}_n^W(x)$ and the
 corresponding coefficients in front of the
operators by $\tilde{K}_n^W$. We list all possible
$\tilde{O}_n^W(x)$ in the Table VI of Appendix A. With these
operators,  (\ref{p6-2}) can also be written as
 \begin{eqnarray}
 S_\mathrm{eff}[U,J]\bigg|_{\mbox{\tiny
anomalous~}p^6}={\displaystyle\sum_{n=1}^{23}}\int
d^4x~\tilde{K}_n^W~\tilde{O}_n^W(x)\;.\label{p6-3}
\end{eqnarray}
Through using the equation of motion, we can obtain the relations among
the two sets of operators $\tilde{O}_n^W(x)$ and $O_n^W(x)$ as follows
\begin{eqnarray}
&&\hspace*{-0.5cm}\tilde{O}_1^W=O_1^W/B_0\hspace*{0.8cm}\tilde{O}_2^W=O_2^W/B_0\hspace*{0.8cm}\tilde{O}_3^W=O_4^W/B_0
\hspace*{0.8cm}\tilde{O}_4^W=O_5^W/B_0\hspace*{0.8cm}\tilde{O}_5^W=O_7^W/B_0\hspace*{0.8cm}\tilde{O}_6^W=O_9^W/B_0\nonumber\\
&&\hspace*{-0.5cm}\tilde{O}_7^W=O_{11}^W/B_0\hspace{0.9cm}\tilde{O}_8^W=O_{12}^W\hspace*{0.9cm}\tilde{O}_9^W=O_1^W
\hspace*{0.9cm}\tilde{O}_{10}^W=O_{16}^W\hspace*{0.9cm}\tilde{O}_{11}^W=O_{17}^W\hspace*{0.9cm}\tilde{O}_{12}^W=O_{13}^W
\hspace*{0.9cm}\tilde{O}_{13}^W=O_{14}^W\nonumber\\
&&\hspace*{-0.5cm}\tilde{O}_{14}^W=O_{15}^W\hspace{0.9cm}\tilde{O}_{15}^W=-O_4^W\!\!+\!\frac{2}{N_f}O_6^W\hspace*{0.9cm}
\tilde{O}_{16}^W=-O_5^W\!\!-\!\frac{1}{N_f}O_6^W
\hspace*{0.9cm}\tilde{O}_{17}^W=O_{19}^W\hspace*{0.9cm}\tilde{O}_{18}^W=O_{20}^W
\hspace*{0.9cm}\tilde{O}_{19}^W=O_{21}^W\nonumber\\
&&\hspace*{-0.5cm}\tilde{O}_{20}^W=O_{22}^W\hspace*{1cm}\tilde{O}_{21}^W=O_{23}^W
 \hspace*{1cm}\tilde{O}_{22}^W=O_7^W\!\!-\!\frac{1}{N_f}O_8^W\hspace*{1cm}\tilde{O}_{23}^W=O_9^W\!\!-\!\frac{1}{N_f}O_{10}^W\;,~~~~
 \label{Nf=3}
\end{eqnarray}
where $B_0$ is the order $p^2$ LEC in the normal part of the chiral
Lagrangian. Here we divide $O^W_1,\cdots,O^W_7$ by $B_0$, making the
 matrices $A_{mn}$ introduced later in
Eq.(\ref{TildeOnExp}) independent of $B_0$. For $N_f=2$,
(\ref{Nf=3}) is changed to
\begin{eqnarray}
&&\hspace*{-0.5cm}\tilde{O}_1^W=0\hspace*{0.7cm}\tilde{O}_2^W=o_1^W/B_0\hspace*{0.7cm}\tilde{O}_3^W=o_2^W/B_0
\hspace*{0.7cm}\tilde{O}_4^W=-o_2^W/(2B_0)\!+\!o_6^W/B_0\hspace*{0.7cm}\tilde{O}_5^W=o_3^W/B_0\hspace*{0.7cm}
\tilde{O}_6^W=o_4^W/B_0\nonumber\\
&&\hspace*{-0.5cm}\tilde{O}_7^W=o_5^W/B_0\hspace{0.8cm}\tilde{O}_8^W=\tilde{O}_9^W=\tilde{O}_{10}^W=
\tilde{O}_{11}^W=0\hspace*{0.8cm}\tilde{O}_{12}^W=-o_9^W
\hspace*{0.8cm}\tilde{O}_{13}^W=\tilde{O}_{14}^W=-\frac{1}{2}o_6^W\!+o_9^W\hspace*{0.8cm}\tilde{O}_{15}^W=-o_6^W\nonumber\\
&&\hspace*{-0.5cm} \tilde{O}_{16}^W=-\frac{1}{2}o_6^W
\hspace*{0.9cm}\tilde{O}_{17}^W=o_{10}^W\hspace*{0.9cm}\tilde{O}_{18}^W=\tilde{O}_{19}^W=-o_{10}^W
\hspace*{0.9cm}\tilde{O}_{20}^W=\frac{1}{4}o_7^W-\frac{1}{8}o_8^W-o^W_{10}+o^W_{11}-2o^W_{13}
\hspace*{0.9cm}\tilde{O}_{21}^W=0
\nonumber\\
&&\hspace*{-0.5cm}\tilde{O}_{22}^W=o_7^W-\frac{1}{2}o_8^W\hspace*{1cm}\tilde{O}_{23}^W=0\;.~~~~
 \label{Nf=2}
\end{eqnarray}
 Direct comparison between (\ref{p6-1}) and
(\ref{p6-3}) is difficult, since in (\ref{p6-1}) we have an extra
 integration over parameter $t$ and the number of operators in (\ref{p6-1}) is much larger than it is in (\ref{p6-3}).
 Instead of finishing the integration over parameter $t$ in (\ref{p6-1}),
 we introduce an integration of parameter $t$ in (\ref{p6-3}).
 Since we are only interested in the  $U$ dependent part of the chiral Lagrangian,
 adding some $U$ field independent pure source terms in (\ref{p6-3}) will not change our
 result; therefore we can rewrite (\ref{p6-3}) as
 \begin{eqnarray}
 S_\mathrm{eff}[U,J]\bigg|_{\mbox{\tiny
anomalous~}p^6}&=&{\displaystyle\sum_{n=1}^{23}}\int
d^4x~\tilde{K}_n^W[\tilde{O}_n^W(x)-\tilde{O}_n^W(x)\bigg|_{U=1}]={\displaystyle\sum_{n=1}^{23}}\int
d^4x~\tilde{K}_n^W[\tilde{O}_n^W(x)\bigg|_{U\rightarrow
U_{t=1}}-\tilde{O}_n^W(x)\bigg|_{U\rightarrow U_{t=0}}]\nonumber\\
&=&{\displaystyle\sum_{n=1}^{23}}\int
d^4x~\tilde{K}_n^W[\tilde{O}_n^W(x)\bigg|_{U\rightarrow
U_t}]\bigg|^{t=1}_{t=0} ={\displaystyle\sum_{n=1}^{23}}\int
d^4x~\tilde{K}_n^W\int_0^1dt~\frac{d}{dt}[\tilde{O}_n^W(x)\bigg|_{U\rightarrow
U_t}]\;.\label{p6-4}
\end{eqnarray}
In expression (\ref{p6-4}), integration of parameter $t$ is already
present in the formula, then the only remaining problem is that in
(\ref{p6-4}) there are only 23 independent terms acted on by the
differential of $t$, while in (\ref{p6-1}) there are 210 terms.
comparing (\ref{p6-4}) and (\ref{p6-1}), we obtain
\begin{eqnarray}
{\displaystyle\sum_{n=1}^{23}}~\tilde{K}_n^W~\frac{d}{dt}[\tilde{O}_n^W(x)\bigg|_{U\rightarrow
U_t}]={\displaystyle\sum_{m=1}^{210}}~\bar{K}_m^W~\bar{O}_m^W(x,t)\;.\label{p6-5}
\end{eqnarray}
 Note that with the help of relation (\ref{difft}),
$\frac{d}{dt}[\tilde{O}_n^W(x)\bigg|_{U\rightarrow U_t}]$ appearing
in the above equation can be reduced to linear composition of
$\bar{O}_m^W(x,t)$, i.e.
\begin{eqnarray}
\frac{d}{dt}[\tilde{O}_n^W(x)\bigg|_{U\rightarrow
U_t}]={\displaystyle\sum_{m=1}^{210}}~A_{nm}\bar{O}_m^W(x,t)\label{TildeOnExp}
\end{eqnarray}
with the $23\times 210$ matrix $A_{nm}$ given by Table VII in
 Appendix B, Then we rearrange (\ref{TildeOnExp}) by multiplying both
sides of the equation by some $23\times 23$ matrix elements
$C_{n'n}$,
\begin{eqnarray}
{\displaystyle\sum_{n=1}^{23}}C_{n'n}\frac{d}{dt}[\tilde{O}_n^W(x)\bigg|_{U\rightarrow
U_t}]={\displaystyle\sum_{n=1}^{23}\sum_{m=1}^{210}}~C_{n'n}A_{nm}\bar{O}_m^W(x,t)
={\displaystyle\sum_{m=1}^{210}}~R_{n'm}\bar{O}_m^W(x,t)\hspace*{1.5cm}R_{n'm}\equiv{\displaystyle\sum_{n=1}^{23}}C_{n'n}A_{nm}
~~~~\label{TildeOnExp1}
\end{eqnarray}
and tune $C_{n'n}$ in such a way that a $23\times 23$ submatrix
$R'$ is a unit matrix, i.e. $R'_{n'm'}=\delta_{n'm'}$ with
$n',m'\!=\!1, 3, 4, 5, 6, 7, 20, 43, 44, 49, 50, 51, 52$, $54, 57,
59, 62, 63, 64, 127, 128, 133, 134$.
The $C$ matrix is found to be of the form $\left(\begin{array}{cc} \bar{C}_{7\times7}&0_{7\times15}\\
0_{15\times7}&\tilde{C}_{15\times15}\end{array}\right)$ where
$\bar{C}$ and $\tilde{C}$ are $7\times7$ and $15\times15$ matrices
respectively. The off diagonal parts are two matrices with null matrix
elements and the dimensions are $7\times15$ and $15\times7$. We
label the dimension of the sub-matrices as their subscripts.
$\bar{C}$ and $\tilde{C}$ matrices are given in Table VIII and
Table IX in Appendix B. We call the remaining part of $R_{n'm}$ the
matrix $R_{n'm''}~m''\neq m'$. Then (\ref{TildeOnExp1}) is changed
to
\begin{eqnarray}
{\displaystyle\sum_{n=1}^{23}}C_{m'n}\frac{d}{dt}[\tilde{O}_n^W(x)\bigg|_{U\rightarrow
U_t}]
=\bar{O}_{m'}^W(x,t)+{\displaystyle\sum_{m''}}~R_{m'm''}\bar{O}_{m''}^W(x,t)\;.
~~~~\label{TildeOnExp2}
\end{eqnarray}
Multiplying both sides of the above equation by $\bar{K}^W_{m'}$,
\begin{eqnarray}
{\displaystyle\sum_{m'}\sum_{n=1}^{23}}\bar{K}^W_{m'}C_{m'n}\frac{d}{dt}[\tilde{O}_n^W(x)\bigg|_{U\rightarrow
U_t}]
={\displaystyle\sum_{m'}}\bar{K}^W_{m'}\bar{O}_{m'}^W(x,t)+{\displaystyle\sum_{m'}\sum_{m''}}~\bar{K}^W_{m'}R_{m'm''}\bar{O}_{m''}^W(x,t)\;.
~~~~\label{TildeOnExp3}
\end{eqnarray}
Comparing (\ref{TildeOnExp3}) and (\ref{p6-5}), to make these
two equations consistent with each other, we must have
conditions,
\begin{eqnarray}
\tilde{K}_n^W={\displaystyle\sum_{m'}}\bar{K}^W_{m'}C_{m'n}\hspace{2cm}\bar{K}^W_{m''}
={\displaystyle\sum_{m'}}~\bar{K}^W_{m'}R_{m'm''}\;,
\end{eqnarray}
 in which the second equation is a consistency check for the
coefficients $\bar{K}^W_{m''}$ of the dependent operators
$\bar{O}_{m''}^W(x,t)$. We have checked analytically that these
constraints are all automatically satisfied and this can be seen as a
consistency check of our formulation. The first equation gives
$\tilde{K}_n^W$ in terms of $\bar{K}^W_{m'}$ and $C_{m'n}$.
Substituting it in the expressions obtained for $\bar{K}_{m'}^W$ and
$C_{m'n}$, we finally obtain the 23 order $p^6$ LECs for the three and
more flavors anomalous part of chiral Lagrangian.

The resulting analytical expressions for $\tilde{K}_n^W$ as functions of
quark self energy $\Sigma$ are given in Appendix C. With
$\tilde{K}_n^W$ given in Appendix C, we can choose a suitable running
coupling constant $\alpha_s(p^2)$, solve the Schwinger-Dyson equation
numerically, obtaining the quark self energy $\Sigma$, and then calculate the
numerical values of all order $p^6$ anomalous LECs. To obtain the
final numerical result, we have assumed $F_0=87$MeV as input to fix
the dimensional parameter $\Lambda_\mathrm{QCD}$ appearing in the
running coupling constant $\alpha_s(p^2)$ and taken momentum cutoff
$\Lambda=1.00^{+0.10}_{-0.10}$GeV. Because of the appearance of the
divergent order $p^2$ LEC $B_0$ in Eqs.(\ref{Nf=3}) and
(\ref{Nf=2}), we need a momentum cutoff $\Lambda$ to make $B_0$
finite as we did previously in Ref.\cite{WQ4}.  In Table I, we give
the numerical values for all 21 nonzero LECs for three
flavors($C_3^W=C_{18}^W=0$ in our formulation).

Combined with our numerical result, we also list the numerical
 estimates for some of the LECs from five different models and different
processes given in
Ref.\cite{p6anomLEC},\cite{p6anomLEC1},\cite{p6anomLEC2},\cite{p6anomLEC3}
and \cite{p6anomLEC4}. In Ref.\cite{p6anomLEC}, model I and III are
all from direct chiral perturbation(ChPT) computations, except that
model I is the full ChPT result, while in model III, low energy
experiment data are extrapolated to the high energy region;~ model II is
the vector meson dominance model (VMD);~ model IV and V are the
chiral constituent quark model (CQM) with some extrapolations
included in model V. For a fixed model, different processes may
give different results. For  example, in model I for $C_7^W$ and models
 I and IV for $C_{22}^W$, we all obtain two results from
two different processes. Further,
Ref.\cite{p6anomLEC},\cite{p6anomLEC3} and \cite{p6anomLEC4} also
give estimations on some combinations or ratios of LECs. We list our
and their results in Table II. For $N_f=2$, in Table III, we give
the numerical values of all 12 nonzero LECs ($c_{12}^W=0$ in our
formulation) which are actually of the very same structure as that
given by \cite{anom1}.
\begin{eqnarray} &&\hspace{-0.5cm}\mbox{\small{\bf TABLE I.}~ The
nonzero values of
the order $p^6$ anomalous LECs $C_1^W,C_2^W,C_4^W,\ldots,C_{17}^W,C_{19}^W,\ldots,C_{23}^W$ for three flavors.}\notag\\
&&\hspace{-0.5cm}\mbox{\small The LECs are in units of
$10^{-3}\mathrm{GeV}^{-2}$.
The 2nd column is our result LECs with the values at $\Lambda=1$GeV with superscript}\notag\\
&&\hspace{-0.5cm}\mbox{\small  the difference caused at
$\Lambda=1.1$GeV (i.e.
$C_i^W\big|_{\Lambda=1.1\mathrm{GeV}}-C_i^W\big|_{\Lambda=1\mathrm{GeV}}$)
and subscript the difference caused at
 $\Lambda=0.9$GeV}\notag\\
 &&\hspace{-0.5cm}\mbox{(i.e.
$C_i^W\big|_{\Lambda=0.9\mathrm{GeV}}-C_i^W\big|_{\Lambda=1\mathrm{GeV}}$).~
The 3rd to 7th columns are results given in
Ref.\cite{p6anomLEC}:~~(I)--ChPT,~~(II)--VMD,}\notag\\
&&\hspace{-0.5cm}\mbox{\small
(III)--ChPT(extrapolation),~~(IV)--CQM,~~(V)--CQM(extrapolation).~The
8th column shows results from
Ref.\cite{p6anomLEC1},\cite{p6anomLEC2},\cite{p6anomLEC3},\cite{p6anomLEC4}.}\notag\\
 &&\hspace{-0.5cm} \begin{array}{|c|c|c|c|c|c|c|c|}
 \hline
 n & C^W_n~\mbox{ours} & \mbox{\cite{p6anomLEC}(I)} & \mbox{\cite{p6anomLEC}(II)}
 & \mbox{\cite{p6anomLEC}(III)}
  & \mbox{\cite{p6anomLEC}(IV)} & \mbox{\cite{p6anomLEC}(V)}&\mbox{\cite{p6anomLEC1},~\cite{p6anomLEC2},~\cite{p6anomLEC3},~\cite{p6anomLEC4}}\\
 \hline
 1 & 4.97^{+0.55}_{-0.79}   & & & & & &\\
 2 & -1.43^{+0.10}_{-0.12}  & -0.32\pm10.4  && 0.78\pm 12.7 & 4.96\pm9.70 & -0.074\pm13.3&\\
 4 & -0.96^{+0.22}_{-0.29}  & 0.28\pm9.19   & &0.67\pm 10.9 & 6.32\pm6.09 & -0.55\pm9.05&\\
 5 & 3.26^{+0.34}_{-0.49}   & 28.50\pm28.83  & &9.38\pm 152.2 & 33.05\pm28.66& 34.51\pm41.13&\\
 6 & 0.91^{+0.03}_{-0.04}   & & & & & &\\
 7 & 1.68^{-0.24}_{+0.31}   & 0.013\pm1.17  & & & 0.51\pm0.06 & &0.1\pm1.2\\
   &                        & 20.3\pm18.7   & & & &&0.1^*\\
 8 & 0.41^{+0.01}_{-0.02}   & 0.76\pm0.18   & & & & &0.58\pm0.20\\
   &                        &   & & & &&0.5^*\\
 9 & 1.15^{-0.03}_{+0.03}  & & & & & &\\
 10 & -0.18^{-0.01}_{+0.01}  & & & & & &\\
 11 & -1.15^{+0.08}_{-0.10} & -6.37\pm4.54  & & & -0.00143\pm0.03& &0.68\pm0.21\\
 12 & -5.13^{-0.15}_{+0.25} & & & & & &\\
 13 & -6.37^{-0.18}_{+0.31} & -74.09\pm55.89& -20.00 & -8.44\pm69.9  & 14.15\pm 15.22 & -7.46\pm19.62&\\
 14 & -2.00^{-0.06}_{+0.10} & 29.99\pm11.14 & -6.01  & 0.72\pm15.3  & 10.23\pm7.56 & -0.58\pm9.77&\\
 15 & 4.17^{+0.12}_{-0.20}  & -25.30\pm 23.93& 2.00   & -3.10\pm28.6 & 19.70\pm7.49 & 8.89\pm9.72&\\
 16 & 3.58^{+0.10}_{-0.17} & & &  & &&\\
 17 & 1.98^{+0.06}_{-0.10}  & & &  & &&\\
 19 & 0.29^{+0.01}_{-0.01} & & &  & &&\\
 20 & 1.83^{+0.05}_{-0.09}  & & &  & &&\\
 21 & 2.48^{+0.07}_{-0.12}  & & &  & &&\\
 22 & 5.01^{+0.14}_{-0.24}  & 6.52\pm0.78   & 8.01 &  & 3.94\pm0.43 &&5.4\pm0.8\\
    &                       & 5.07\pm0.71   & & & 3.94 \pm0.43 &&\\
 23 & 2.74^{+0.08}_{-0.13}  & & &  & &&\\
 \hline
 \end{array}\nonumber
 \end{eqnarray}
 $^*$~This result is just the absolute value given in Ref.\cite{p6anomLEC3}.
 \begin{eqnarray} &&\hspace{-0.5cm}\mbox{\small{\bf TABLE II.}~ Some combinations or ratios of LECs
in units of $10^{-3}\mathrm{GeV}^{-2}$.
The 2nd column is our result LECs}\notag\\
&&\hspace{-0.5cm}\mbox{\small  with the values at $\Lambda=1$GeV, and
with superscript the difference caused at
$\Lambda=1.1$GeV and subscript the difference }\notag\\
&&\hspace{-0.5cm}\mbox{\small caused at
 $\Lambda=0.9$GeV.~ The 3rd to 5th columns are results given in
Ref.\cite{p6anomLEC}:~~(I)--ChPT,~~(II)--VMD,~(III)--ChPT}\notag\\
&&\hspace{-0.5cm}\mbox{\small~(extrapolation),~~(IV)--CQM,~~(V)--CQM
(extrapolation). The 6th and 7th columns are results given in
Ref.\cite{p6anomLEC3}}\notag\\
&&\hspace{-0.5cm}\mbox{\small and \cite{p6anomLEC4} respectively. }\notag\\
&&\hspace{-0.5cm}\begin{array}{|c|c|c|c|c|c|c|}
 \hline & \mbox{ours} & \mbox{\cite{p6anomLEC}} & \mbox{\cite{p6anomLEC}}& \mbox{\cite{p6anomLEC}}&\mbox{\cite{p6anomLEC3}}&\mbox{\cite{p6anomLEC4}}\\
 \hline
 C^W_3-C^W_6                  & -0.91^{-0.03}_{+0.04} & 21.67\pm17.41~\mbox{(I)} & 5.07\pm5.07~\mbox{(IV)}&-2.14\pm6.54~\mbox{(V)}&&\\
 2C^W_{15}-4C^W_{14}+C^W_{13} & 9.95^{+0.29}_{-0.48} & -244.7\pm148.4~\mbox{(I)} & \approx8.0~\mbox{(II)} &-17.52\pm188.3~\mbox{(III)}&&\\
 2C^W_{14}-C^W_{13}           & 2.38^{+0.07}_{-0.12}  & 134.1\pm78.17~\mbox{(I)} & \approx8.0~\mbox{(II)} &9.88\pm100.5~\mbox{(III)}&&\\
|C_7^W|/|C_8^W| & 4.12^{-0.69}_{+1.01}& &&&0.2&<0.1\\
 \hline
 \end{array}\notag
 \end{eqnarray}
 \begin{eqnarray} &&\hspace{-0.5cm}\mbox{\small{\bf TABLE III.}~
The nonzero values of the $p^6$ order anomalous LECs
$c_1^W,\ldots,c_{11}^W,c_{13}^W$ for two flavor in units of
$10^{-3}\mathrm{GeV}^{-2}$.}\notag\\
&&\hspace{-1cm}{\footnotesize\begin{array}{|c|c|c|c|c|c|c|c|c|c|c|c|}
 \hline c_1^W & c_2^W & c_3^W& c_4^W& c_5^W& c_6^W& c_7^W& c_8^W& c_9^W& c_{10}^W& c_{11}^W& c_{13}^W\\
\hline-1.46^{+0.10}_{-0.12}& -1.25^{+0.09}_{-0.11}&
2.96^{-0.20}_{+0.25} & 0.63^{-0.04}_{+0.05}& -1.17^{+0.08}_{-0.10}&
0.77^{+0.26}_{-0.36} & -0.04^{-0.00}_{+0.00} & 0.02^{+0.00}_{-0.00}
& 8.19^{+0.23}_{-0.38}
 & -8.73^{-0.24}_{+0.41} & 4.85^{+0.13}_{-0.23}& -9.70^{-0.27}_{+0.45}  \\
  \hline
 \end{array}}\notag
 \end{eqnarray}
 We see that most of our results are consistent with
those we have found in the literature.

As a phenomenological check for two flavor anomalous LECs, we
discuss the $\pi^0\rightarrow\gamma\gamma$ process.
Ref.\cite{p6anomLEC4} gives the amplitude of this process by
 \begin{eqnarray}
 T_\mathrm{LO+NLO}&=&\frac{1}{F}\bigg\{\frac{1}{4\pi^2}+\frac{16}{3}m_{\pi}^2(-4c_3^{Wr}-4c_7^{Wr}+c_{11}^{Wr})
 +\frac{64}{9}B(m_d-m_u)(5c_3^{Wr}+c_r^{Wr}+2c_8^{Wr})\bigg\}\;.\label{amp}
 \end{eqnarray}
In our calculation, we choose the center value
 $B(m_d-m_u)=0.32m^2_{\pi^0}$ given in Ref.\cite{p6anomLEC4}. Experimentally, the $\pi^0\rightarrow\gamma\gamma$ process
dominates the life time of $\pi^0$ to $98.79\%$, and if we ignore
that small fraction from other processes, then the life time of
$\pi^0$ can be expressed in terms of amplitude $T$ as
$1/\tau=\pi\alpha m_\pi^3T^2/4$. In Table IV we give our result for
$\tau_\mathrm{LO}$ up to the leading order $p^4$, which corresponds
to the first term of the r.h.s of Eq.({\ref{amp}), and
$\tau_\mathrm{NLO}$ up to the next leading order $p^6$ of the low
energy expansion. Experimental result from particle data
group\cite{PDG} is also included in the table for comparison.
\begin{eqnarray}
&&\hspace{-0.5cm}\mbox{\small{\bf TABLE IV.}~ $\pi^0$ life time in
units of $10^{-17}\mathrm{s}$.}\notag\\
&& \begin{array}{|c|c|c|}
 \hline
               & \tau_{\rm LO} & \tau_{\rm NLO}\\
 \hline
 F=87\mathrm{MeV} & 7.56 &  7.59^{-0.03}_{+0.04}\\
 F=93\mathrm{MeV}       & 8.63 &  8.67^{-0.03}_{+0.04}\\
 \hline
 \mbox{Exp.\cite{PDG}}          & \multicolumn{2}{c}{8.4\pm0.6}\vline                    \\
 \hline
 \end{array}\notag
 \end{eqnarray}
Our result roughly matches the experimental value and we see that the order $p^6$ results have less effect
 on the life time of $\pi^0$.
\section{Summary and Future Work}

In this work, we review the general anomaly structure of the
effective chiral Lagrangian and then generalize our order $p^6$
calculation in Ref.\cite{WQ4} from the normal part to the anomalous
part of the chiral Lagrangian for pseudoscalar mesons. The result is obtained by
computing the imaginary $\Sigma$ dependent part of
Tr$\ln[\slashed{\partial}+J_{\Omega}+\Sigma(-\bar{\nabla}^2)]$. To
match the calculation of the order $p^4$ anomalous part, in practice
we calculate the integration of parameter $t$ over
$\frac{d}{dt}\mathrm{Tr}\ln[i\slashed{\partial}+J_{\Omega(t)}+\Sigma(-\nabla_t^2)]$.
The conventional chiral Lagrangian is also reformulated to an integration of $t$ and through comparison of it with our
result, we read out all order $p^6$ anomalous LECs expressed in
terms of quark self energy $\Sigma$. Inputting the SDE solution of
$\Sigma(k^2)$, we obtain numerical values and compare them with
those we can find in literature. Some of them are consistent, some
are not. We leave those inconsistent results to future
investigations. Combined with the previous result on the order $p^6$
normal LECs given in Ref.\cite{WQ4}, we have now completed
all the order $p^6$ LECs computations. Based on them, one direction
of the further research is to apply the order $p^6$ chiral
Lagrangian to various pseudoscalar meson processes and discuss the
corresponding physics. Another direction is to improve the precision
of (\ref{Seff}) and our ladder approximation SDE. With these
improvements, we expect a more precise estimation on all LECs in
future.
\section*{Acknowledgments}
This work was  supported by National  Science Foundation of China
(NSFC) under Grant No.10875065.


\newpage
\appendix
\section{List of All Operators $\bar{O}^{\mu\nu\lambda\rho}_n$ and $\tilde{O}^W_n$}

In this appendix, we first explicitly write down all 210
$\bar{O}^{\mu\nu\lambda\rho}_n$ operators. To save the space, we use
some simplified symbols to represent the original symbols in the text.
Our $\bar{O}^{\mu\nu\lambda\rho}_n$s are constructed in such a way
that they are invariant under charge conjugation transformation.
This causes the result that most of $\bar{O}^{\mu\nu\lambda\rho}_n$s
consist of two terms which are charge conjugates to each other.
\begin{eqnarray}
&&\hspace{-0.5cm}\mbox{\small\bf TABLE
V.}~~~~~~~\mu\equiv\nabla_t^\mu,~~\nu\equiv\nabla_t^\nu,~~\lambda\equiv\nabla_t^\lambda,~~
\rho\equiv\nabla_t^\rho,~~\bar{\mu}\equiv a_t^\mu,~~\bar{\nu}\equiv
a_t^\nu,~~\bar{\lambda}\equiv a_t^\lambda,~~ \bar{\rho}\equiv
a_t^\rho,~~s\equiv s_t,~~p\equiv p_t\notag\\
&&\hspace{-0.5cm}\cdots\sigma\cdots\sigma\equiv\cdots\nabla_t^\sigma\cdots\nabla_{t,\sigma},
~~\cdots\bar{\sigma}\cdots\bar{\sigma}\equiv\cdots a_t^\sigma\cdots
a_{t,\sigma}
,~~\cdots\sigma\cdots\bar{\sigma}\equiv\cdots\nabla_t^\sigma\cdots
a_{t,\sigma},~~\cdots\bar{\sigma}\cdots\sigma\equiv\cdots
a_t^\sigma\cdots\nabla_{t,\sigma}\notag\\
 &&\hspace{0.5cm} {\scriptsize
\begin{array}{|c|c|c|c|c|c|c|c|c|c|}
 \hline n & \bar{O}^{\mu\nu\lambda\rho}_n &n & \bar{O}^{\mu\nu\lambda\rho}_n &n & \bar{O}^{\mu\nu\lambda\rho}_n &n & \bar{O}^{\mu\nu\lambda\rho}_n&n & \bar{O}^{\mu\nu\lambda\rho}_n \\
 \hline
 1& s \mu \nu \lambda \rho  +\mu \nu \lambda \rho s  & 43& \mu \nu \lambda \rho \sigma \sigma  +\sigma \sigma \mu \nu \lambda \rho  & 85& \mu \sigma \bar{\nu}\lambda \rho \bar{\sigma} +\bar{\sigma}\mu \nu \bar{\lambda}\sigma \rho  & 127& \mu \nu \bar{\lambda}\bar{\rho}\bar{\sigma}\bar{\sigma} +\bar{\sigma}\bar{\sigma}\bar{\mu}\bar{\nu}\lambda \rho  & 169& \mu \bar{\nu}\bar{\lambda}\bar{\sigma}\bar{\sigma}\rho  +\mu \bar{\sigma}\bar{\sigma}\bar{\nu}\bar{\lambda}\rho  \\
 2& \mu s \nu \lambda \rho  +\mu \nu \lambda s \rho  & 44& \mu \nu \lambda \sigma \rho \sigma  +\sigma \mu \sigma \nu \lambda \rho  & 86& \mu \sigma \bar{\nu}\lambda \sigma \bar{\rho} +\bar{\mu}\sigma \nu \bar{\lambda}\sigma \rho  & 128& \mu \nu \bar{\lambda}\bar{\sigma}\bar{\rho}\bar{\sigma} +\bar{\sigma}\bar{\mu}\bar{\sigma}\bar{\nu}\lambda \rho  & 170& \mu \bar{\nu}\bar{\sigma}\bar{\lambda}\bar{\rho}\sigma  +\sigma \bar{\mu}\bar{\nu}\bar{\sigma}\bar{\lambda}\rho  \\
 3& \mu \nu s \lambda \rho  & 45& \mu \nu \lambda \sigma \sigma \rho  +\mu \sigma \sigma \nu \lambda \rho  & 87& \mu \sigma \bar{\nu}\sigma \lambda \bar{\rho} +\bar{\mu}\nu \sigma \bar{\lambda}\sigma \rho  & 129& \mu \nu \bar{\lambda}\bar{\sigma}\bar{\sigma}\bar{\rho} +\bar{\mu}\bar{\sigma}\bar{\sigma}\bar{\nu}\lambda \rho  & 171& \mu \bar{\nu}\bar{\sigma}\bar{\lambda}\bar{\sigma}\rho  +\mu \bar{\sigma}\bar{\nu}\bar{\sigma}\bar{\lambda}\rho  \\
\hline
 \end{array}}\notag
 \end{eqnarray}
\newpage
Next, we list all 23 $\otw_n$ operators.
 \begin{eqnarray}
 &&\hspace{-0.5cm}\mbox{\small{\bf TABLE
VI.}~~~~~List of $\otw_n$ operators, where we divide
$\tilde{O}^W_1,...,\tilde{O}^W_7$ by $B_0$ making the matrices
$A_{mn}$ introduced}\notag\\
&&\hspace{-0.5cm}\mbox{in Eq.(36) independent of $B_0$.~~ The
symbols are introduced in Ref.\cite{p6-1}.~The comparisons between}\notag\\
&&\hspace{-0.5cm}\mbox{\small  the symbols introduced in
Ref.\cite{p6-1} and ours are given in Table
XV. of Ref.\cite{WQ4}.}\notag\\
 &&\hspace{-0.5cm}
\begin{array}{|c|c|c|c|}
 \hline n & \otw_n &n & \otw_n \\
 \hline 1& \la i u^{\mu}u^{\nu}u^{\lambda}u^{\rho}\chim\ra \epsilon_{\mu\nu\lambda\rho}/B_0 &
 13& -i \la \fp^{\mu\nu}u_{\sigma}u^{\lambda}h^{\rho\sigma}\ra\epsilon_{\mu\nu\lambda\rho}
 -i \la \fp^{\mu\nu}h^{\lambda\sigma}u^{\rho}u_{\sigma}\ra\epsilon_{\mu\nu\lambda\rho}\\
 \hline 2& \la u^{\mu}u^{\nu}\chip \fm^{\lambda\rho}\ra\epsilon_{\mu\nu\lambda\rho}/B_0
 - \la u^{\mu}u^{\nu}\fm^{\lambda\rho}\chip\ra \epsilon_{\mu\nu\lambda\rho}/B_0&
 14& -i \la \fp^{\mu\nu}u^{\lambda}h^{\rho\sigma}u_{\sigma}\ra\epsilon_{\mu\nu\lambda\rho}
 -i \la \fp^{\mu\nu}u_{\sigma}h^{\lambda\sigma}u^{\rho}\ra\epsilon_{\mu\nu\lambda\rho}\\
 \hline 3& \la \fp^{\mu\nu}u^{\lambda}u^{\rho}\chim\ra \epsilon_{\mu\nu\lambda\rho}/B_0
 + \la \fp^{\mu\nu}\chim u^{\lambda}u^{\rho}\ra\epsilon_{\mu\nu\lambda\rho}/B_0&
 15& i \la \fp^{\mu\nu}u^{\lambda}u^{\rho}h^{\sigma}_{~\sigma}\ra\epsilon_{\mu\nu\lambda\rho}
 +i \la \fp^{\mu\nu}h^{\sigma}_{~\sigma}u^{\lambda}u^{\rho}\ra\epsilon_{\mu\nu\lambda\rho}\\
 \hline 4& - \la \fp^{\mu\nu}u^{\lambda}\chim u^{\rho}\ra\epsilon_{\mu\nu\lambda\rho}/B_0&
 16& -i \la \fp^{\mu\nu}u^{\lambda}h^{\sigma}_{~\sigma}u^{\rho}\ra\epsilon_{\mu\nu\lambda\rho}\\
 \hline 5& i \la \fp^{\mu\nu}\fp^{\lambda\rho}\chim\ra \epsilon_{\mu\nu\lambda\rho}/B_0&
 17& i \la {\fp^{\mu}}_{\sigma}u^{\nu}u^{\lambda}\fm^{\rho\sigma}\ra\epsilon_{\mu\nu\lambda\rho}
 +i \la {\fp^{\mu}}_{\sigma}\fm^{\nu\sigma}u^{\lambda}u^{\rho}\ra\epsilon_{\mu\nu\lambda\rho}\\
 \hline 6& i \la \fm^{\mu\nu}\fm^{\lambda\rho}\chim\ra \epsilon_{\mu\nu\lambda\rho}/B_0&
18& i \la
\fp^{\mu\sigma}u^{\nu}u_{\sigma}\fm^{\lambda\rho}\ra\epsilon_{\mu\nu\lambda\rho}
 -i \la \fp^{\mu\sigma}\fm^{\nu\lambda}u_{\sigma}u^{\rho}\ra\epsilon_{\mu\nu\lambda\rho} \\
\hline
 \end{array}\notag
 \end{eqnarray}
\newpage\section{$A$, $C$ and $R$ matrices}

In this appendix, we give matrix $A_{nm}$, $C_{m'n}$ and $R_{m'm}$.
For convenience of writing, in practice, we do not write $A$ matrix,
but its transverse $A^T$ multiplied by $-i$.
 \begin{eqnarray}
 &&\hspace{4.5cm}\mbox{\small\bf TABLE
VII.}~~~~~~~-i(A^T)_{mn}~\mbox{matrix}\notag\\
&&{\scriptsize\begin{array}{|c|c|c|c|c|c|c|c|c|c|c|c|c|c|c|c|c|c|c|c|c|c|c|c|}
 \hline
 m,n  & 1 & 2 & 3 & 4 & 5 & 6 & 7 & 8 & 9 & 10 & 11 & 12 & 13 & 14 & 15 & 16 & 17 & 18 & 19 & 20 & 21 & 22 & 23 \\
 \hline
 1 & 0  & 0  & 0  & 0  & 32  & 0  & 32  & 0  & 0  & 0  & 0  & 0  & 0  & 0  & 0  & 0  & 0  & 0  & 0  & 0  & 0  & 0  & 0  \\
 2 & 0  & 0  & 0  & 0  & 0  & 0  & 0  & 0  & 0  & 0  & 0  & 0  & 0  & 0  & 0  & 0  & 0  & 0  & 0  & 0  & 0  & 0  & 0  \\
 3 & 0  & 0  & 0  & 0  & 0  & 0  & -64  & 0  & 0  & 0  & 0  & 0  & 0  & 0  & 0  & 0  & 0  & 0  & 0  & 0  & 0  & 0  & 0  \\
 4 & 0  & 0  & -32  & 0  & -32  & 0  & -32  & 0  & 0  & 0  & 0  & 0  & 0  & 0  & 0  & 0  & 0  & 0  & 0  & 0  & 0  & 0  & 0  \\
 5 & 0  & 0  & 0  & 0  & 0  & -32  & 32  & 0  & 0  & 0  & 0  & 0  & 0  & 0  & 0  & 0  & 0  & 0  & 0  & 0  & 0  & 0  & 0  \\
 6 & 0  & -32  & 0  & 0  & 0  & -32  & 32  & 0  & 0  & 0  & 0  & 0  & 0  & 0  & 0  & 0  & 0  & 0  & 0  & 0  & 0  & 0  & 0  \\
 7 & 0  & 0  & 0  & -32  & 0  & -32  & 32  & 0  & 0  & 0  & 0  & 0  & 0  & 0  & 0  & 0  & 0  & 0  & 0  & 0  & 0  & 0  & 0  \\
\hline
 \end{array}}\notag
 \end{eqnarray}
\newpage
 \begin{eqnarray}
  &&\hspace{3.5cm}\mbox{\small\bf TABLE
VII~(continued).}~~~~~~~-i(A^T)_{mn}~\mbox{matrix}\notag\\
 &&{\scriptsize\begin{array}{|c|c|c|c|c|c|c|c|c|c|c|c|c|c|c|c|c|c|c|c|c|c|c|c|}
 \hline
 m,n  & 1 & 2 & 3 & 4 & 5 & 6 & 7 & 8 & 9 & 10 & 11 & 12 & 13 & 14 & 15 & 16 & 17 & 18 & 19 & 20 & 21 & 22 & 23 \\
 \hline
56 & 0  & 0  & 0  & 0  & 0  & 0  & 0  & 0  & 0  & 0  & 0  & 8  & -48  & 0  & -32  & 0  & -8  & 0  & 0  & 0  & 0  & 32  & 0  \\
 57 & 0  & 0  & 0  & 0  & 0  & 0  & 0  & 0  & 0  & 0  & 0  & 24  & -16  & 0  & 32  & 0  & 8  & 0  & 0  & 16  & 0  & -32  & 0  \\
 58 & 0  & 0  & 0  & 0  & 0  & 0  & 0  & 0  & 0  & 0  & 0  & -16  & 0  & 0  & -32  & 0  & 0  & 0  & 0  & -8  & 0  & 32  & 0  \\
 59 & 0  & 0  & 0  & 0  & 0  & 0  & 0  & 0  & 0  & 0  & 0  & 0  & -16  & 32  & 0  & 32  & 0  & 0  & 0  & 0  & 0  & 0  & -32  \\
 60 & 0  & 0  & 0  & 0  & 0  & 0  & 0  & 0  & 0  & 0  & 0  & 0  & -16  & 0  & 0  & 0  & 0  & 0  & 0  & 8  & 0  & 0  & 0  \\
61 & 0  & 0  & 0  & 0  & 0  & 0  & 0  & 0  & 0  & 0  & 0  & 0  & 32  & 0  & 32  & 0  & 0  & 0  & 0  & -8  & 8  & 0  & 0  \\
  62 & 0  & 0  & 0  & 0  & 0  & 0  & 0  & 0  & 0  & 0  & 0  & 0  & 0  & 16  & 0  & 0  & 0  & -16  & -16  & -24  & 16  & 0  & 0  \\
\hline
 \end{array}}\notag
 \end{eqnarray}
\newpage
 \begin{eqnarray}
   &&\hspace{4.5cm}\mbox{\small\bf TABLE
VII~(continued).}~~~~~~~-i(A^T)_{mn}~\mbox{matrix}\notag\\
 &&{\scriptsize\begin{array}{|c|c|c|c|c|c|c|c|c|c|c|c|c|c|c|c|c|c|c|c|c|c|c|c|}
 \hline
 m,n  & 1 & 2 & 3 & 4 & 5 & 6 & 7 & 8 & 9 & 10 & 11 & 12 & 13 & 14 & 15 & 16 & 17 & 18 & 19 & 20 & 21 & 22 & 23 \\
 \hline
 115 & 0  & 0  & 0  & 0  & 0  & 0  & 0  & 0  & 0  & 0  & 0  & 0  & -16  & -32  & -32  & -32  & 0  & 0  & 0  & -8  & -8  & 32  & 32  \\
 116 & 0  & 0  & 0  & 0  & 0  & 0  & 0  & 0  & 0  & 0  & 0  & -8  & 0  & 0  & 0  & 32  & -8  & 16  & 0  & 8  & 24  & 0  & -32  \\
 117 & 0  & 0  & 0  & 0  & 0  & 0  & 0  & 0  & 0  & 0  & 0  & -8  & 16  & 0  & 0  & 0  & 8  & 0  & 0  & 0  & -16  & 0  & 0  \\
 118 & 0  & 0  & 0  & 0  & 0  & 0  & 0  & 0  & 0  & 0  & 0  & 8  & 0  & 0  & 0  & -32  & 8  & -16  & 0  & -8  & 8  & 0  & 32  \\
 119 & 0  & 0  & 0  & 0  & 0  & 0  & 0  & 0  & 0  & 0  & 0  & 8  & -16  & 0  & 0  & 0  & -8  & 0  & 0  & 0  & 0  & 0  & 0  \\
 120 & 0  & 0  & 0  & 0  & 0  & 0  & 0  & 0  & 0  & 0  & 0  & 0  & -32  & 32  & 0  & 32  & 0  & 0  & 0  & 0  & 0  & 0  & -32  \\
 121 & 0  & 0  & 0  & 0  & 0  & 0  & 0  & 0  & 0  & 0  & 0  & 0  & 0  & 0  & 0  & 0  & 0  & 0  & 0  & 8  & 0  & -64  & 0  \\
\hline
 \end{array}}\notag
 \end{eqnarray}
\newpage
 \begin{eqnarray}
   &&\hspace{4.5cm}\mbox{\small\bf TABLE
VII~(continued).}~~~~~~~-i(A^T)_{mn}~\mbox{matrix}\notag\\
 &&{\scriptsize\begin{array}{|c|c|c|c|c|c|c|c|c|c|c|c|c|c|c|c|c|c|c|c|c|c|c|c|}
 \hline
 m,n  & 1 & 2 & 3 & 4 & 5 & 6 & 7 & 8 & 9 & 10 & 11 & 12 & 13 & 14 & 15 & 16 & 17 & 18 & 19 & 20 & 21 & 22 & 23 \\
 \hline
  174 & 0  & 0  & 0  & 0  & 0  & 0  & 0  & 0  & 0  & 0  & 0  & -8  & 80  & 0  & 64  & 0  & 8  & 0  & 0  & 16  & 8  & -64  & 0  \\
 175 & 0  & 0  & 0  & 0  & 0  & 0  & 0  & 0  & 0  & 0  & 0  & 8  & -32  & 0  & 0  & 0  & -8  & 16  & 0  & -8  & 0  & 0  & 0  \\
 176 & 0  & 0  & 0  & 0  & 0  & 0  & 0  & 0  & 0  & 0  & 0  & -40  & 16  & -16  & -64  & 0  & -8  & 0  & -16  & -40  & 0  & 64  & 0  \\
 177 & 0  & 0  & 0  & 0  & 0  & 0  & 0  & 0  & 0  & 0  & 0  & 24  & 0  & -16  & 0  & 0  & 8  & -16  & 16  & 48  & -8  & 0  & 0  \\
 178 & 0  & 0  & 0  & 0  & 0  & 0  & 0  & 0  & 0  & 0  & 0  & -32  & 32  & 0  & -32  & 0  & 0  & 0  & 0  & -40  & 8  & 32  & -32  \\
 179 & 0  & 0  & 0  & 0  & 0  & 0  & 0  & 0  & 0  & 0  & 0  & 32  & 0  & 16  & 64  & 0  & 0  & 0  & 16  & 24  & 0  & -64  & -32  \\
 180 & 0  & 0  & 0  & 0  & 0  & 0  & 0  & 0  & 0  & 0  & 0  & -16  & 0  & 16  & 0  & 0  & 0  & 0  & -16  & -24  & 0  & 0  & 32  \\
\hline
 \end{array}}\notag
 \end{eqnarray}
 From (\ref{TildeOnExp1}), $C$ matrix is consist of two sub-matrices
 $\bar{C}$ and $\tilde{C}$.  $\bar{C}$ matrix is
  \begin{eqnarray}
&&\hspace{1cm}\mbox{\small\bf TABLE.
VIII}~~~~~~~\bar{C}_{m'n}~\mbox{matrix}\notag\\
 &&\begin{array}{|c|c|c|c|c|c|c|c|}
 \hline
 m',n  & 1 & 2 & 3 & 4 & 5 & 6 & 7 \\
 \hline
 1 & -\frac{i}{32}  & 0  & -\frac{i}{32}  & 0  & \frac{i}{32}  & 0  & 0  \\
 3 & 0  & 0  & 0  & 0  & \frac{i}{64}  & -\frac{i}{64}  & -\frac{i}{64}  \\
 4 & -\frac{i}{16}  & 0  & -\frac{i}{32}  & 0  & 0  & 0  & 0  \\
 5 & 0  & \frac{i}{32}  & 0  & \frac{i}{32}  & 0  & -\frac{i}{32}  & 0  \\
 6 & \frac{i}{32}  & -\frac{i}{32}  & 0  & 0  & 0  & 0  & 0  \\
 7 & -\frac{i}{32}  & 0  & 0  & -\frac{i}{32}  & 0  & 0  & 0  \\
 20 & -\frac{i}{32}  & 0  & 0  & 0  & 0  & 0  & 0  \\
 \hline
 \end{array}\notag
 \end{eqnarray}
\newpage$\tilde{C}$ matrix is
\begin{eqnarray}
&&\hspace{4cm}\mbox{\small\bf TABLE.
IX}~~~~~~~\tilde{C}_{m'n}~\mbox{matrix}\notag\\
 &&\begin{array}{|r|r|r|r|r|r|r|r|r|r|r|r|r|r|r|r|r|}
 \hline
 m',n  & 8 & 9 & 10 & 11 & 12 & 13 & 14 & 15 & 16 & 17 & 18 & 19 & 20 & 21 & 22 & 23 \\
 \hline
 43 & \frac{19i}{160}  & -\frac{11i}{80}  & -\frac{19i}{160}  & -\frac{13i}{160}  & -\frac{3i}{80}  & -\frac{i}{80}  & -\frac{7i}{160}  & \frac{9i}{160}  & -\frac{i}{160}  & -\frac{13i}{80}  & -\frac{3i}{40}  & -\frac{i}{32}  & -\frac{i}{40}  & -\frac{i}{10}  & -\frac{3i}{160}  & -\frac{7i}{160}  \\
 44 & \frac{31i}{320}  & -\frac{3i}{20}  & -\frac{21i}{320}  & -\frac{17i}{320}  & \frac{3i}{160}  & \frac{i}{160}
 & -\frac{13i}{320}  & \frac{11i}{320}  & \frac{11i}{320}  & -\frac{17i}{160}  & -\frac{9i}{160}  & \frac{i}{64}
 & -\frac{i}{20}  & -\frac{3i}{40}  & -\frac{i}{160}  & -\frac{3i}{320}  \\
 49 & -\frac{i}{16}  & \frac{3i}{32}  & \frac{3i}{32}  & \frac{i}{16}  & \frac{i}{16}  & \frac{i}{32}  & \frac{i}{32}
 & -\frac{i}{16}  & 0  & \frac{i}{8}  & \frac{i}{16}  & \frac{i}{32}  & 0  & \frac{i}{16}  & 0  & \frac{i}{64}  \\
 50 & 0  & 0  & 0  & 0  & 0  & 0  & -\frac{i}{32}  & 0  & \frac{i}{32}  & 0  & -\frac{i}{16}  & \frac{i}{32}  & 0  & 0
 & 0  & 0  \\
 51 & -\frac{5i}{64}  & \frac{3i}{32}  & \frac{5i}{64}  & \frac{3i}{64}  & \frac{i}{32}  & 0  & \frac{i}{64}
 & -\frac{3i}{64}  & \frac{i}{64}  & \frac{3i}{32}  & \frac{i}{32}  & \frac{3i}{64}  & 0  & \frac{i}{16}  & 0
 & \frac{i}{32}  \\
 52 & \frac{5i}{32}  & -\frac{3i}{16}  & -\frac{5i}{32}  & -\frac{3i}{32}  & 0  & 0  & -\frac{i}{16}  & \frac{i}{16}
 & \frac{i}{32}  & -\frac{i}{4}  & -\frac{i}{8}  & -\frac{i}{16}  & 0  & -\frac{i}{8}  & 0  & -\frac{i}{32}  \\
 54 & -\frac{3i}{32}  & \frac{i}{32}  & \frac{5i}{32}  & \frac{i}{32}  & 0  & 0  & \frac{i}{16}  & -\frac{i}{32}  & 0
 & \frac{i}{8}  & \frac{i}{16}  & \frac{i}{8}  & 0  & \frac{i}{8}  & 0  & \frac{i}{16}  \\
\hline
 \end{array}\notag
 \end{eqnarray}
\section{Final Analytical Result on $\tilde{K}^W_n$}

In this appendix, we list our analytical result on 23 LECs for $p^6$
order anomalous part of the chiral Lagrangian,
\begin{eqnarray}
\tilde{K}^W_1&=&\int\frac{d^4k}{(2\pi)^4}\bigg[-\frac{1}{2}k^2\sk^3
X^5 -\frac{1}{2} \sk^5 X^5 +\frac{3}{4} k^4 \sk^2 \skp X^5 +\frac{3}{4} k^2 \sk^4 \skp X^5 \bigg]\notag\\
 \tilde{K}^W_2&=&\int\frac{d^4k}{(2\pi)^4}\bigg[-\frac{1}{4} \sk^5 X^5
 +\frac{1}{8} k^4 \sk^2 \skp X^5 +\frac{5}{8} k^2 \sk^4 \skp X^5 \bigg]\notag\\
 \tilde{K}^W_3&=&\int\frac{d^4k}{(2\pi)^4}\bigg[-\frac{1}{4} k^2 \sk^3 X^5
 -\frac{1}{4} \sk^5 X^5+\frac{1}{2} k^4 \sk^2 \skp X^5+\frac{1}{2} k^2 \sk^4 \skp X^5\bigg]\notag\\
 \tilde{K}^W_4&=&\int\frac{d^4k}{(2\pi)^4}\bigg[-\frac{1}{4} k^2 \sk^3 X^5
 -\frac{1}{4} \sk^5 X^5+\frac{1}{2} k^4 \sk^2 \skp X^5+\frac{1}{2} k^2 \sk^4 \skp X^5\bigg]\notag\\
 \tilde{K}^W_5&=&\int\frac{d^4k}{(2\pi)^4}\bigg[\frac{3}{16} k^2 \sk^3 X^5
 +\frac{3}{16} \sk^5 X^5-\frac{1}{16} k^6 \skp X^5-\frac{1}{2} k^4 \sk^2 \skp X^5-\frac{7}{16} k^2 \sk^4 \skp X^5\bigg]\notag\\
 \tilde{K}^W_6&=&\int\frac{d^4k}{(2\pi)^4}\bigg[\frac{1}{16} k^2 \sk^3 X^5+\frac{1}{16} \sk^5 X^5
 -\frac{1}{8} k^4 \sk^2 \skp X^5 -\frac{1}{8} k^2 \sk^4 \skp X^5]\notag\\
 \tilde{K}^W_7&=&\int\frac{d^4k}{(2\pi)^4}\bigg[-\frac{1}{16} k^2 \sk^3 X^5-\frac{1}{16} \sk^5 X^5
 +\frac{1}{32} k^6 \skp X^5 +\frac{3}{16} k^4 \sk^2 \skp X^5+\frac{5}{32} k^2 \sk^4 \skp X^5\bigg]\notag\\
 \tilde{K}^W_{8}&=&\int\frac{d^4k}{(2\pi)^4}\bigg[(\frac{9}{40} k^2\skpp -\frac{1}{40}\sk^2 \skpp
 +\frac{3}{40} k^4  \skppp  +\frac{1}{180} k^2 \sk^2 \skppp )k^2\sk X^4+(-\frac{29}{80} k^4 \sk \skp
 +\frac{17}{80} k^2 \sk^3 \skp +\frac{3}{40} \sk^5 \skp
  \notag\\
 &&+\frac{7}{16} k^6 \skp^2
 -\frac{3}{4} k^4 \sk^2 \skp^2 -\frac{3}{16} k^2 \sk^4 \skp^2 -\frac{67}{240} k^6 \sk \skpp
 +\frac{31}{120} k^4 \sk^3 \skpp+\frac{3}{80} k^2 \sk^5 \skpp  +\frac{27}{80} k^8 \skp \skpp-\frac{1}{2} k^6 \sk^2 \skp \skpp
 \notag\\
 &&  +\frac{13}{80} k^4 \sk^4 \skp \skpp) X^5+(-\frac{17}{80} k^4 \sk^2 -\frac{1}{8} k^2 \sk^4
-\frac{3}{80} \sk^6  +\frac{151}{240} k^6 \sk \skp-\frac{5}{8} k^4
\sk^3 \skp  +\frac{13}{80} k^2 \sk^5 \skp-\frac{59}{120} k^8 \skp^2
 \notag\\
 && +\frac{893}{480} k^6 \sk^2 \skp^2
  -\frac{217}{240} k^4 \sk^4 \skp^2  +\frac{39}{160} k^2 \sk^6 \skp^2  -\frac{293}{240} k^8 \sk \skp^3
   +\frac{5}{8} k^6 \sk^3 \skp^3 -\frac{39}{80} k^4 \sk^5 \skp^3) X^6
 \bigg]\notag\\
 \tilde{K}^W_9&=&\int\frac{d^4k}{(2\pi)^4}\bigg[(-\frac{7}{40} k^2\skpp +\frac{3}{40}  \sk^2 \skpp -\frac{7}{120} k^4 \skppp
 -\frac{1}{360} k^2 \sk^2 \skppp)k^2\sk X^4 +(\frac{37}{80} k^4 \sk \skp -\frac{1}{80} k^2 \sk^3 \skp
 +\frac{1}{40} \sk^5 \skp \notag\\
 && -\frac{11}{16} k^6 \skp^2+\frac{1}{4} k^4 \sk^2 \skp^2  -\frac{1}{16} k^2 \sk^4 \skp^2
 +\frac{27}{80} k^6 \sk \skpp  -\frac{1}{15} k^4 \sk^3 \skpp+\frac{1}{80} k^2 \sk^5 \skpp -\frac{8}{15} k^8 \skp \skpp
 -\frac{1}{6} k^6 \sk^2 \skp \skpp   \notag\\
 && -\frac{7}{15} k^4 \sk^4 \skp \skpp) X^5 +(\frac{8}{15} k^4 \sk^2  +\frac{1}{4} k^2 \sk^4
 +\frac{3}{10} \sk^6  -\frac{139}{120} k^6 \sk \skp+\frac{7}{12} k^4 \sk^3 \skp
 -\frac{17}{40} k^2 \sk^5 \skp +\frac{17}{20} k^8 \skp^2\notag\\
 &&-\frac{317}{240} k^6 \sk^2 \skp^2+\frac{23}{120} k^4 \sk^4 \skp^2 -\frac{51}{80} k^2 \sk^6 \skp^2
 +\frac{197}{120} k^8 \sk \skp^3 +\frac{11}{12} k^6 \sk^3 \skp^3 +\frac{51}{40} k^4 \sk^5 \skp^3) X^6
 \bigg]\notag\\
 \tilde{K}^W_{10}&=&\int\frac{d^4k}{(2\pi)^4}\bigg[(-\frac{33}{80} k^2 \skpp +\frac{27}{80} \sk^2 \skpp
 -\frac{11}{80} k^4\skppp  +\frac{71}{720} k^2 \sk^2 \skppp)k^2\sk X^4
 +(\frac{17}{40} k^4 \sk \skp  -\frac{31}{40} k^2 \sk^3 \skp  +\frac{3}{10} \sk^5 \skp
 \notag\\
 && -\frac{1}{4} k^6 \skp^2 +2 k^4 \sk^2 \skp^2 -\frac{3}{4} k^2 \sk^4 \skp^2 +\frac{107}{240} k^6 \sk \skpp
 -\frac{217}{240} k^4 \sk^3 \skpp +\frac{3}{20} k^2 \sk^5 \skpp  -\frac{7}{30} k^8 \skp \skpp
 +\frac{23}{12} k^6 \sk^2 \skp \skpp \notag\\
 && -\frac{17}{20} k^4 \sk^4 \skp \skpp)X^5 +(\frac{11}{240} k^4 \sk^2  -\frac{7}{8} k^2 \sk^4 +\frac{43}{80} \sk^6
 -\frac{37}{80} k^6 \sk \skp+\frac{71}{24} k^4 \sk^3 \skp  -\frac{213}{80} k^2 \sk^5 \skp
 +\frac{29}{120} k^8 \skp^2 \notag\\
 &&  -\frac{641}{160} k^6 \sk^2 \skp^2 +\frac{1127}{240} k^4 \sk^4 \skp^2
 -\frac{89}{160} k^2 \sk^6 \skp^2  +\frac{283}{240} k^8 \sk \skp^3
 -\frac{97}{24} k^6 \sk^3 \skp^3+\frac{89}{80} k^4 \sk^5 \skp^3) X^6
 \bigg]\notag\\
 \tilde{K}^W_{11}&=&\int\frac{d^4k}{(2\pi)^4}\bigg[(-\frac{17}{160} k^2 \skpp +\frac{3}{160} \sk^2 \skpp
 -\frac{7}{180} k^4 \skppp +\frac{1}{360} k^2 \sk^2 \skppp)k^2\sk X^4
 +(\frac{9}{40} k^4 \sk \skp -\frac{1}{20} k^2 \sk^3 \skp -\frac{1}{40} \sk^5 \skp\notag\\
 &&-\frac{5}{16} k^6 \skp^2 +\frac{1}{4} k^4 \sk^2 \skp^2+\frac{1}{16} k^2 \sk^4 \skp^2 +\frac{49}{240} k^6 \sk \skpp
 -\frac{7}{120} k^4 \sk^3 \skpp-\frac{1}{80} k^2 \sk^5 \skpp  -\frac{31}{120} k^8 \skp \skpp+\frac{1}{8} k^6 \sk^2 \skp \skpp\notag\\
 && -\frac{7}{60} k^4 \sk^4 \skp \skpp )X^5+(\frac{9}{80} k^4 \sk^2+\frac{1}{80} \sk^6-\frac{3}{10} k^6 \sk \skp
 +\frac{1}{2} k^4 \sk^3 \skp +\frac{1}{20} k^2 \sk^5 \skp +\frac{2}{5} k^8 \skp^2 -\frac{541}{480} k^6 \sk^2 \skp^2\notag\\
 && -\frac{41}{240} k^4 \sk^4 \skp^2-\frac{23}{160} k^2 \sk^6 \skp^2+\frac{221}{240} k^8 \sk \skp^3
 +\frac{5}{24} k^6 \sk^3 \skp^3 +\frac{23}{80} k^4 \sk^5 \skp^3) X^6\bigg]\notag\\
 \tilde{K}^W_{12}&=&\int\frac{d^4k}{(2\pi)^4}\bigg[(-\frac{1}{40} k^2\skpp-\frac{1}{40} \sk^2 \skpp -\frac{1}{120} k^4  \skppp
 +\frac{7}{360} k^2 \sk^2 \skppp)k^2\sk X^4+(-\frac{1}{20} k^4 \sk \skp-\frac{1}{10} k^2 \sk^3 \skp -\frac{1}{20} \sk^5 \skp
 \notag\\
 && +\frac{1}{8} k^6 \skp^2 X^5+\frac{1}{4} k^4 \sk^2 \skp^2 +\frac{1}{8} k^2 \sk^4 \skp^2  -\frac{1}{120} k^6 \sk \skpp
 -\frac{1}{5} k^4 \sk^3 \skpp -\frac{1}{40} k^2 \sk^5 \skpp +\frac{7}{80} k^8 \skp \skpp +\frac{7}{24} k^6 \sk^2 \skp \skpp
\notag\\
 &&-\frac{31}{240} k^4 \sk^4 \skp \skpp)X^5+(-\frac{53}{120} k^4 \sk^2 -\frac{1}{2} k^2 \sk^4  +\frac{1}{40} \sk^6
 +\frac{11}{15} k^6 \sk \skp +\frac{5}{6} k^4 \sk^3 \skp -\frac{2}{5} k^2 \sk^5 \skp -\frac{7}{60} k^8 \skp^2\notag\\
 &&-\frac{37}{80} k^6 \sk^2 \skp^2 +\frac{59}{120} k^4 \sk^4 \skp^2 -\frac{13}{80} k^2 \sk^6 \skp^2 -\frac{3}{40} k^8 \sk \skp^3-\frac{5}{12} k^6 \sk^3 \skp^3
 +\frac{13}{40} k^4 \sk^5 \skp^3) X^6\bigg]\notag\\
 \tilde{K}^W_{13}&=&\int\frac{d^4k}{(2\pi)^4}\bigg[(\frac{1}{80} k^2 \skpp +\frac{1}{80}\sk^2 \skpp +\frac{1}{240} k^4 \skppp
 +\frac{1}{240} k^2 \sk^2 \skppp)k^2\sk X^4+(-\frac{1}{80} k^2 \sk^3 \skp-\frac{3}{80} k^4 \sk \skp
  +\frac{1}{40} \sk^5 \skp\notag\\
 &&+\frac{1}{16} k^6 \skp^2 -\frac{1}{16} k^2 \sk^4 \skp^2 -\frac{1}{60} k^6 \sk \skpp -\frac{1}{240} k^4 \sk^3 \skpp
 +\frac{1}{80} k^2 \sk^5 \skpp +\frac{7}{240} k^8 \skp \skpp -\frac{7}{240} k^4 \sk^4 \skp \skpp) X^5\notag\\
 &&+(-\frac{7}{40} k^4 \sk^2 -\frac{1}{8} k^2 \sk^4+\frac{1}{20} \sk^6 +\frac{41}{120} k^6 \sk \skp
 +\frac{1}{6} k^4 \sk^3 \skp -\frac{7}{40} k^2 \sk^5 \skp
 -\frac{1}{15} k^8 \skp^2 +\frac{13}{240} k^6 \sk^2 \skp^2\notag\\
 &&+\frac{13}{120} k^4 \sk^4 \skp^2  -\frac{1}{80} k^2 \sk^6 \skp^2 -\frac{13}{120} k^8 \sk \skp^3 -\frac{1}{12} k^6 \sk^3 \skp^3 +\frac{1}{40} k^4 \sk^5 \skp^3) X^6
 \bigg]\notag\\
 \tilde{K}^W_{14}&=&\int\frac{d^4k}{(2\pi)^4}\bigg[(-\frac{9}{80} k^2 \skpp+\frac{11}{80}\sk^2 \skpp -\frac{2}{45} k^4 \skppp
   +\frac{1}{40}k^2 \sk^2 \skppp)k^2\sk X^4 +(\frac{7}{80} k^4 \sk \skp-\frac{21}{80} k^2 \sk^3 \skp
   +\frac{3}{20} \sk^5 \skp\notag\\
 && +\frac{5}{8} k^4 \sk^2 \skp^2-\frac{3}{8} k^2 \sk^4 \skp^2 +\frac{13}{120} k^6 \sk \skpp -\frac{19}{60} k^4 \sk^3 \skpp
 +\frac{3}{40} k^2 \sk^5 \skpp-\frac{1}{30} k^8 \skp \skpp +\frac{13}{24} k^6 \sk^2 \skp \skpp
\notag\\
 &&  -\frac{17}{40} k^4 \sk^4 \skp \skpp) X^5+(\frac{13}{240} k^4 \sk^2 +\frac{19}{80} \sk^6
 +\frac{1}{120} k^6 \sk \skp  +\frac{7}{12} k^4 \sk^3 \skp
 -\frac{47}{40} k^2 \sk^5 \skp -\frac{1}{40} k^8 \skp^2-\frac{113}{160} k^6 \sk^2 \skp^2\notag\\
 &&+\frac{197}{80} k^4 \sk^4 \skp^2-\frac{57}{160} k^2 \sk^6 \skp^2 -\frac{7}{80} k^8 \sk \skp^3 -\frac{41}{24} k^6 \sk^3 \skp^3
 +\frac{57}{80} k^4 \sk^5 \skp^3) X^6 \bigg]\notag\\
 \tilde{K}^W_{15}&=&\int\frac{d^4k}{(2\pi)^4}\bigg[(\frac{11}{160} k^2\skpp-\frac{9}{160} \sk^2 \skpp +\frac{19}{720} k^4 \skppp
 -\frac{11}{720} k^2 \sk^2 \skppp)k^2\sk X^4 +(-\frac{9}{80} k^4 \sk \skp +\frac{7}{80} k^2 \sk^3 \skp\notag\\
 && -\frac{1}{20} \sk^5 \skp +\frac{1}{8} k^6 \skp^2 -\frac{1}{4} k^4 \sk^2 \skp^2 +\frac{1}{8} k^2 \sk^4 \skp^2 -\frac{2}{15} k^6 \sk \skpp
 +\frac{11}{120} k^4 \sk^3 \skpp -\frac{1}{40} k^2 \sk^5 \skpp +\frac{19}{160} k^8 \skp \skpp\notag\\
 &&-\frac{3}{16} k^6 \sk^2 \skp \skpp +\frac{31}{160} k^4 \sk^4 \skp \skpp) X^5+(\frac{3}{80} k^4 \sk^2
 -\frac{13}{80} \sk^6 +\frac{1}{40} k^6 \sk \skp -\frac{1}{4} k^4 \sk^3 \skp+\frac{19}{40} k^2 \sk^5 \skp\notag\\
 && -\frac{1}{5} k^8 \skp^2 +\frac{343}{480} k^6 \sk^2 \skp^2 -\frac{97}{240} k^4 \sk^4 \skp^2
 +\frac{29}{160} k^2 \sk^6 \skp^2 -\frac{103}{240} k^8 \sk \skp^3 +\frac{5}{24} k^6 \sk^3 \skp^3
 -\frac{29}{80} k^4 \sk^5 \skp^3)X^6 \bigg]\notag\\
 \tilde{K}^W_{16}&=&\int\frac{d^4k}{(2\pi)^4}\bigg[(-\frac{7}{80} k^2\skpp -\frac{7}{80}\sk^2 \skpp -\frac{1}{45} k^4\skppp
 -\frac{1}{120} k^2 \sk^2 \skppp)k^2\sk X^4+(\frac{3}{40} k^4 \sk \skp -\frac{1}{10} k^2 \sk^3 \skp
 -\frac{7}{40} \sk^5 \skp\notag\\
 && -\frac{1}{16} k^6 \skp^2 +\frac{3}{8} k^4 \sk^2 \skp^2 +\frac{7}{16} k^2 \sk^4 \skp^2 +\frac{23}{240} k^6 \sk \skpp
 +\frac{1}{120} k^4 \sk^3 \skpp -\frac{7}{80} k^2 \sk^5 \skpp -\frac{3}{80} k^8 \skp \skpp
 +\frac{5}{12} k^6 \sk^2 \skp \skpp\notag\\
 &&+\frac{109}{240} k^4 \sk^4 \skp \skpp)X^5+( -\frac{61}{240} k^4 \sk^2-\frac{3}{8} k^2 \sk^4 -\frac{13}{80} \sk^6
 +\frac{23}{120} k^6 \sk \skp +\frac{2}{3} k^4 \sk^3 \skp+\frac{29}{40} k^2 \sk^5 \skp +\frac{1}{20} k^8 \skp^2\notag\\
 && -\frac{209}{160} k^6 \sk^2 \skp^2 -\frac{109}{80} k^4 \sk^4 \skp^2 +\frac{79}{160} k^2 \sk^6 \skp^2
 +\frac{49}{80} k^8 \sk \skp^3 -\frac{1}{24} k^6 \sk^3 \skp^3 -\frac{79}{80} k^4 \sk^5 \skp^3) X^6
 \bigg]\notag\\
 \tilde{K}^W_{17}&=&\int\frac{d^4k}{(2\pi)^4}\bigg[(-\frac{2}{5} k^2\skpp +\frac{1}{10}\sk^2 \skpp -\frac{2}{15} k^4\skppp
 +\frac{11}{180} k^2 \sk^2 \skppp)k^2\sk X^4+(\frac{23}{40} k^4 \sk \skp -\frac{19}{40} k^2 \sk^3 \skp
 -\frac{1}{20} \sk^5 \skp\notag\\
 && -\frac{5}{8} k^6 \skp^2 +\frac{3}{2} k^4 \sk^2 \skp^2 +\frac{1}{8} k^2 \sk^4 \skp^2 +\frac{8}{15} k^6 \sk \skpp
 -\frac{79}{120} k^4 \sk^3 \skpp -\frac{1}{40} k^2 \sk^5 \skpp -\frac{119}{240} k^8 \skp \skpp\notag\\
 && +\frac{29}{24} k^6 \sk^2 \skp \skpp -\frac{151}{240} k^4 \sk^4 \skp \skpp) X^5+(\frac{37}{120} k^4 \sk^2
 -\frac{1}{2} k^2 \sk^4 -\frac{9}{40} \sk^6  -\frac{14}{15} k^6 \sk \skp +\frac{8}{3} k^4 \sk^3 \skp
 +\frac{1}{10} k^2 \sk^5 \skp\notag\\
 &&+\frac{43}{60} k^8 \skp^2 -\frac{1031}{240} k^6 \sk^2 \skp^2 +\frac{53}{40} k^4 \sk^4 \skp^2
  -\frac{53}{80} k^2 \sk^6 \skp^2 +\frac{271}{120} k^8 \sk \skp^3 -\frac{13}{12} k^6 \sk^3 \skp^3
+\frac{53}{40} k^4 \sk^5 \skp^3)X^6 \bigg]\notag\\
 \tilde{K}^W_{18}&=&\int\frac{d^4k}{(2\pi)^4}\bigg[(-\frac{9}{80} k^2\skpp +\frac{11}{80}\sk^2 \skpp-\frac{2}{45} k^4\skppp
 +\frac{7}{180} k^2 \sk^2 \skppp)k^2\sk X^4+(\frac{3}{20} k^4 \sk \skp-\frac{1}{5} k^2 \sk^3 \skp
 +\frac{3}{20} \sk^5 \skp\notag\\
 &&-\frac{1}{8} k^6 \skp^2 +\frac{1}{2} k^4 \sk^2 \skp^2 -\frac{3}{8} k^2 \sk^4 \skp^2 +\frac{13}{120} k^6 \sk \skpp
 -\frac{19}{60} k^4 \sk^3 \skpp +\frac{3}{40} k^2 \sk^5 \skpp -\frac{7}{60} k^8 \skp \skpp
 +\frac{1}{4} k^6 \sk^2 \skp \skpp\notag\\
 && -\frac{19}{30} k^4 \sk^4 \skp \skpp )X^5+(-\frac{1}{20} k^4 \sk^2 -\frac{1}{8} k^2 \sk^4+\frac{7}{40} \sk^6
 +\frac{2}{15} k^6 \sk \skp +\frac{7}{12} k^4 \sk^3 \skp -\frac{21}{20} k^2 \sk^5 \skp +\frac{1}{10} k^8 \skp^2\notag\\
 && -\frac{97}{240} k^6 \sk^2 \skp^2 +\frac{223}{120} k^4 \sk^4 \skp^2 -\frac{51}{80} k^2 \sk^6 \skp^2
 +\frac{17}{120} k^8 \sk \skp^3 -\frac{7}{12} k^6 \sk^3 \skp^3+\frac{51}{40} k^4 \sk^5 \skp^3) X^6
 \bigg]\notag\\
 \tilde{K}^W_{19}&=&\int\frac{d^4k}{(2\pi)^4}\bigg[(-\frac{1}{4} k^2\skpp+\frac{1}{4}\sk^2 \skpp -\frac{1}{12} k^4\skppp
 +\frac{5}{72} k^2 \sk^2 \skppp)k^2\sk X^4+(\frac{3}{16} k^4 \sk \skp -\frac{9}{16} k^2 \sk^3 \skp+\frac{1}{4} \sk^5 \skp\notag\\
 && +\frac{11}{8} k^4 \sk^2 \skp^2 -\frac{5}{8} k^2 \sk^4 \skp^2+\frac{1}{4} k^6 \sk \skpp-\frac{5}{8} k^4 \sk^3 \skpp
 +\frac{1}{8} k^2 \sk^5 \skpp-\frac{1}{24} k^8 \skp \skpp  +\frac{37}{24} k^6 \sk^2 \skp \skpp\notag\\
 &&-\frac{5}{12} k^4 \sk^4 \skp \skpp) X^5+(-\frac{1}{48} k^4 \sk^2 -\frac{1}{4} k^2 \sk^4 +\frac{5}{16} \sk^6-\frac{1}{6} k^6 \sk \skp
 +\frac{4}{3} k^4 \sk^3 \skp-\frac{7}{4} k^2 \sk^5 \skp -\frac{1}{24} k^8 \skp^2\notag\\
 &&-\frac{91}{32} k^6 \sk^2 \skp^2+\frac{173}{48} k^4 \sk^4 \skp^2-\frac{3}{32} k^2 \sk^6 \skp^2 +\frac{25}{48} k^8 \sk \skp^3 -\frac{29}{8} k^6 \sk^3 \skp^3
 +\frac{3}{16} k^4 \sk^5 \skp^3) X^6 \bigg]\notag\\
 \tilde{K}^W_{20}&=&\int\frac{d^4k}{(2\pi)^4}\bigg[(\frac{1}{40} k^2\skpp+\frac{1}{40}\sk^2 \skpp -\frac{1}{180} k^4\skppp
 -\frac{1}{180} k^2 \sk^2 \skppp)k^2\sk X^4+(\frac{1}{20} k^4 \sk \skp +\frac{1}{10} k^2 \sk^3 \skp
 +\frac{1}{20} \sk^5 \skp\notag\\
 &&-\frac{1}{8} k^6 \skp^2 -\frac{1}{4} k^4 \sk^2 \skp^2 -\frac{1}{8} k^2 \sk^4 \skp^2 +\frac{11}{120} k^6 \sk \skpp
   +\frac{7}{60} k^4 \sk^3 \skpp +\frac{1}{40} k^2 \sk^5 \skpp -\frac{13}{120} k^8 \skp \skpp\notag\\
 &&-\frac{1}{12} k^6 \sk^2 \skp \skpp+\frac{1}{40} k^4 \sk^4 \skp \skpp) X^5+(\frac{3}{20} k^4 \sk^2 +\frac{1}{4} k^2 \sk^4 +\frac{1}{10} \sk^6
 -\frac{2}{5} k^6 \sk \skp -\frac{1}{2} k^4 \sk^3 \skp-\frac{1}{10} k^2 \sk^5 \skp\notag\\
 && +\frac{1}{5} k^8 \skp^2 -\frac{1}{10} k^6 \sk^2 \skp^2-\frac{1}{5} k^4 \sk^4 \skp^2 +\frac{1}{10} k^2 \sk^6 \skp^2
 +\frac{1}{5} k^8 \sk \skp^3 -\frac{1}{5} k^4 \sk^5 \skp^3) X^6 \bigg]\notag\\
 \tilde{K}^W_{21}&=&\int\frac{d^4k}{(2\pi)^4}\bigg[(-\frac{11}{40} k^2\skpp+\frac{9}{40}\sk^2 \skpp -\frac{11}{120} k^4\skppp
 +\frac{3}{40} k^2 \sk^2 \skppp)k^2\sk X^4+(\frac{13}{40} k^4 \sk \skp -\frac{19}{40} k^2 \sk^3 \skp
 +\frac{1}{5} \sk^5 \skp\notag\\
 &&-\frac{1}{4} k^6 \skp^2 +\frac{5}{4} k^4 \sk^2 \skp^2-\frac{1}{2} k^2 \sk^4 \skp^2 +\frac{13}{40} k^6 \sk \skpp
 -\frac{23}{40} k^4 \sk^3 \skpp+\frac{1}{10} k^2 \sk^5 \skpp -\frac{11}{60} k^8 \skp \skpp
 +\frac{5}{4} k^6 \sk^2 \skp \skpp\notag\\
 && -\frac{17}{30} k^4 \sk^4 \skp \skpp) X^5+(\frac{1}{10} k^4 \sk^2-\frac{1}{4} k^2 \sk^4+\frac{3}{20} \sk^6
 -\frac{31}{60} k^6 \sk \skp +\frac{4}{3} k^4 \sk^3 \skp -\frac{23}{20} k^2 \sk^5 \skp  +\frac{3}{10} k^8 \skp^2\notag\\
 && -\frac{169}{60} k^6 \sk^2 \skp^2+\frac{38}{15} k^4 \sk^4 \skp^2-\frac{7}{20} k^2 \sk^6 \skp^2
 +\frac{29}{30} k^8 \sk \skp^3 -\frac{7}{3} k^6 \sk^3 \skp^3+\frac{7}{10} k^4 \sk^5 \skp^3) X^6\bigg]\notag\\
 \tilde{K}^W_{22}&=&\int\frac{d^4k}{(2\pi)^4}\bigg[(-\frac{1}{80} k^2\skpp -\frac{1}{80}\sk^2 \skpp -\frac{1}{240} k^4\skppp
 -\frac{1}{240} k^2 \sk^2 \skppp)k^2\sk X^4+(\frac{3}{80} k^4 \sk \skp+\frac{1}{80} k^2 \sk^3 \skp-\frac{1}{40} \sk^5 \skp
 \notag\\
 &&-\frac{1}{16} k^6 \skp^2+\frac{1}{16} k^2 \sk^4 \skp^2+\frac{3}{80} k^6 \sk \skpp+\frac{1}{40} k^4 \sk^3 \skpp
 -\frac{1}{80} k^2 \sk^5 \skpp-\frac{19}{480} k^8 \skp \skpp+\frac{1}{48} k^6 \sk^2 \skp \skpp\notag\\
 && +\frac{29}{480} k^4 \sk^4 \skp \skpp) X^5+(-\frac{1}{80} k^4 \sk^2 -\frac{1}{16} k^2 \sk^4-\frac{1}{20} \sk^6
 +\frac{1}{80} k^6 \sk \skp +\frac{1}{4} k^4 \sk^3 \skp+\frac{19}{80} k^2 \sk^5 \skp+\frac{1}{40} k^8 \skp^2\notag\\
 && -\frac{13}{40} k^6 \sk^2 \skp^2 -\frac{11}{40} k^4 \sk^4 \skp^2+\frac{3}{40} k^2 \sk^6 \skp^2
 +\frac{3}{20} k^8 \sk \skp^3 -\frac{3}{20} k^4 \sk^5 \skp^3)X^6  \bigg]\notag\\
 \tilde{K}^W_{23}&=&\int\frac{d^4k}{(2\pi)^4}\bigg[(-\frac{23}{160}k^2\skpp +\frac{17}{160}\sk^2 \skpp -\frac{23}{480} k^4\skppp
 +\frac{17}{480} k^2 \sk^2 \skppp)k^2\sk X^4+(\frac{19}{160} k^4 \sk \skp-\frac{47}{160} k^2 \sk^3 \skp
 +\frac{7}{80} \sk^5 \skp\notag\\
 &&-\frac{1}{32} k^6 \skp^2+\frac{3}{4} k^4 \sk^2 \skp^2-\frac{7}{32} k^2 \sk^4 \skp^2+\frac{31}{240} k^6 \sk \skpp
 -\frac{157}{480} k^4 \sk^3 \skpp +\frac{7}{160} k^2 \sk^5 \skpp -\frac{11}{480} k^8 \skp \skpp
 +\frac{19}{24} k^6 \sk^2 \skp \skpp\notag\\
 &&-\frac{89}{480} k^4 \sk^4 \skp \skpp) X^5+(-\frac{9}{80} k^4 \sk^2-\frac{3}{16} k^2 \sk^4+\frac{7}{40} \sk^6
 +\frac{1}{120} k^6 \sk \skp+\frac{7}{12} k^4 \sk^3 \skp-\frac{37}{40} k^2 \sk^5 \skp+\frac{1}{60} k^8 \skp^2\notag\\
 && -\frac{223}{160} k^6 \sk^2 \skp^2+\frac{371}{240} k^4 \sk^4 \skp^2-\frac{7}{160} k^2 \sk^6 \skp^2
 +\frac{23}{80} k^8 \sk \skp^3 -\frac{13}{8} k^6 \sk^3 \skp^3+\frac{7}{80} k^4 \sk^5 \skp^3) X^6\bigg]
 \end{eqnarray}
 where $B_0$ is the LEC appear in $p^2$ order normal part of the
 chiral Lagrangian. $\Sigma_k\equiv\Sigma(k^2)$ and
 $X\equiv\frac{1}{k^2+\sk^2}$.
\end{document}